\documentclass [12pt]{article}
\usepackage{latexsym}
\usepackage{amsfonts}
\usepackage{amsmath}
\usepackage{acronym}
\newcommand{\be}{\begin{equation}}
\newcommand{\ee}{\end{equation}}
\newcommand{\bea}{\begin{eqnarray}}
\newcommand{\eea}{\end{eqnarray}}
\makeatletter
\@addtoreset{equation}{section}
\def\theequation{\thesection.\arabic{equation}}
 \makeatother
\begin{document}
\normalsize
\title{\Large Nontrivial topological dynamics in Minkowskian Higgs model quantized by Dirac.}
\author
{{\bf L.~D.~Lantsman}\\
 Wissenschaftliche Gesellschaft bei
 J$\rm \ddot u$dische Gemeinde  zu Rostock,\\Augusten Strasse, 20,\\
 18055, Rostock, Germany; \\ 
Tel.  049-0381-799-07-24,\\
llantsman@freenet.de}
\maketitle
\begin {abstract}
We study the nontrivial topological dynamics inherent in the Minkowskian Higgs model with vacuum BPS monopoles quantized by Dirac.

It comes to persistent collective solid rotations inside the physical BPS monopole vacuum, accompanied by never vanishing vacuum "electric" fields (vacuum monopoles) $\bf E$.

The enumerated rotary effects inside the physical BPS monopole vacuum suffered the Dirac fundamental quantization are the specific display of the Josephson effect, whose nature will be reveal in the present study.
\end{abstract}
 \newpage
\tableofcontents
\newpage
\section{Introduction.}
Continuing the investigations about the Dirac fundamental quantization \cite{Dir} of the Minkowskian Higgs model with vacuum BPS monopole solutions started in the recent papers \cite{rem1, rem2}, now we concentrate our efforts to the nontrivial topological dynamics, proves to be inherent in the enumerated model side by side with manifest superfluid properties \cite{rem2}.

\medskip
The origin of that nontrivial topological dynamics (we reveal in the present study) lies in the Gauss-shell reduction of the Minkowskian Higgs model (with vacuum BPS monopole solutions).

At resolving the Yang-Mills (YM) Gauss law constraint
\be
\label{Gauss}
\frac {\delta W}{\delta A^a_0}=0 \Longleftrightarrow [D^2(A)]^{ac}A_{0c}= D^{ac}_i(A)\partial_0 A_{c}^i
\ee
with the covariant (Coulomb) gauge \cite{LP2,LP1}
$$A^{a\parallel}\sim [D^{ac}_i(\Phi ^{(0)})A_c^{i }]=0\vert _{t=0}
$$
(in the fixed time instant $t_0$ and in the [topologically trivial] YM BPS monopole background $\Phi ^{(0)}$),
the former turns into the second-order homogeneous differential equation
\be \label{homo}   [D^2_i(\Phi ^{(0)})]^{ac} A_{0c}=0,        \ee
permitting the family of so-called {\it zero mode   solutions} \cite{Pervush1, Pervush2} 
\be \label{zero}  A_0^c(t,{\bf x})= {\dot N}(t) \Phi_{(0)}^c ({\bf x})\equiv Z^c,
   \ee
implicating the topological variable $\dot N(t)$ and Higgs (topologically trivial) vacuum Higgs BPS monopole modes $\Phi_0^a ({\bf x})$.

\medskip
The appearance of $Z^a$ solutions at the constraint-shell reduction of the Minkowskian Higgs model with vacuum BPS monopoles involves lot of important consequences for this  model, we shall discuss in the present study.

First of all, knowing YM potentials $Z^a$ (referring obviously to the BPS monopole vacuum), it is easy to  write down $F_{i0}^a$ components of the YM tension tensor, taking the shape of so-called vacuum "electric" monopoles \cite{LP2,LP1}
$$   F^a_{i0}={\dot N}(t)D ^{ac}_i(\Phi_k ^{(0)})\Phi_{0c}({\bf x}).         $$
Issuing from vacuum "electric" monopoles $F_{i0}^a$, one can construct \cite{LP2,LP1, David2, fund} the action functional
\be \label{rot} W_N=\int d^4x \frac {1}{2}(F_{0i}^c)^2 =\int dt\frac {{\dot N}^2 I}{2}\ee
involving the  rotary momentum \cite{David2}
\be \label{I} I=\int \sb {V} d^3x (D_i^{ac}(\Phi_k^{(0)})\Phi_{0c})^2 =
\frac {4\pi^2\epsilon}{ \alpha _s}
=\frac {4\pi^2}{\alpha _s^2}\frac {1}{ V<B^2>}.    \ee
The YM coupling constant 
$$\alpha _s=\frac{g^2}{4\pi (\hbar c)^2 } $$
enters this expression for $I$ together with the typical size $\epsilon $ \cite{LP2,LP1}
 of BPS monopoles and the vacuum expectation value $<B^2>$ of the "magnetic" field.

It will be shown that the action functional $ W_N $ describes correctly collective solid rotations of the BPS monopole vacuum (suffered the Dirac fundamental quantization \cite{Dir}) with constant angular velocities ${\dot N}(t)$.  

These  constant angular velocities ${\dot N}(t)$ determine the real energy-momentum spectrum
\be \label{pin} P_N ={\dot N} I= 2\pi k +\theta; \quad \theta  \in [-\pi,\pi];      \ee
of the free rotator $W_N$,  accompanied by the wave function
\be  \label{psin}  \Psi _N\equiv <P_N\vert N>=\exp (iP_N N).            \ee 
\medskip

The enumerated rotary effects inside the   BPS monopole vacuum suffered the Dirac fundamental quantization \cite{Dir} may be explained good \cite{Pervush3}  as a particular manifestation of the {\it Josephson effect}, coming to persistent circular motions of material points (in particular, quantum fields)  without (outward) sources. \par
\medskip
In  the present study we consider in detail vacuum rotary effects in the Minkowskian Higgs model with vacuum BPS monopoles quantized by Dirac, associated, in the first place, with properties of the topological variable $N(t)$, proving to be the noninteger degree of the map referring to the $U(1)\to SU(2)$ embedding. \par
All this implies the following plan of the article, we now propose our readers. \par
\medskip 
{\it Section 2}, devoted to the nontrivial topological dynamics inherent in the Minkowskian Higgs model with vacuum BPS monopoles quantized by Dirac \cite{Dir}, consists of two subsections. \par
In {\it Subsection 2.1}  we study more in detail how the nontrivial topological dynamics arises  in the Minkowskian Higgs model quantized by Dirac and involving vacuum BPS monopole solutions. \par
Repeating the arguments \cite{LP2,LP1,Pervush2}, we show that its origin in the constraint-shell reduction of this model in terms of {\it topological Dirac variables} $\hat A^D_i$ ($i=1,2$) \cite{rem2,David2}: gauge invariant and transverse functionals of YM fields, satisfying the covariant transverse gauge. \par
\medskip 
Although  Subsection 2.1 contains, actually, no new information in comparison with Refs. \cite{LP2,LP1,Pervush2}, it will be helpful, nevertheless, as the preparatory material for  {\it Subsection 2.2}, in which we  study the properties of the  topological variable $N(t)$, the noninteger degree of the map referring to the $U(1)\to SU(2)$ embedding. \par
The principal result will be got in Subsection 2.2 is that \cite{Pervush3}
\be \label{Poi} {\dot N}(t)={\rm const} = (n_{\rm out}-n_{\rm in})/T\equiv \nu/T,        \ee 
where $n_{\rm out}$, $n_{\rm in}$~ $\in {\bf Z}$ refer to the fixed time instants $t=\pm T/2$, respectively; $\nu$ is referred to as the {\it Pontryagin index } \cite{Pervush2, Pervush3}.
\par
It will be argued herewith (due to general QFT reasoning) that it would be set $T\to\infty$.\par
By that the claim to $N(t)$ to take integers at $t=\pm T/2\to\infty$ is satisfied automatically when one represents this variable as \cite{Pervush3} 
\be \label {tbo}  N(t)=(n_+ - n_-)\frac t T + (n_+ + n_-) \frac 1 2;
      \ee
$$ n_-\equiv n_{\rm in}; \quad n_+\equiv n_{\rm out}. $$
\medskip
{\it Section 3} will be devoted to  the general discussion about the Josephson effect, in which the arguments \cite{Pervush3} will be repeated. \par
It will be shown that persistent circular motions of  quantum fields (that is the essence of the  Josephson effect) are characterized \cite{fund,Pervush3} by never vanishing (until $\theta  \neq 0$) momenta
$$ P=   \hbar ~\frac {2\pi k+ \theta} L, $$
with $L$ being the length of the whole closed line along which the given quantum field moves. \par
Such momenta $P$ attain their nonzero minima 
$ p=\hbar \theta /L $ 
as $k=0$ and if $\theta  \neq 0$. \par
Besides collective solid rotations inside the BPS monopole vacuum suffered the Dirac fundamental quantization \cite{Dir} (these rotations are characterized constant angular velocities $\dot N(t)$), the Josephson effect in the appropriate Minkowskian Higgs model comes to never vanishing (until $\theta  \neq 0$) vacuum "electric" fields ("electric" monopoles)
\be
\label{sem}   (E_i^a)_{\rm min}= \theta \frac {\alpha_s}{4\pi^2\epsilon} B_i^a; \quad -\pi\leq \theta \leq \pi.
      \ee
Such minimum value of the vacuum "electric" field $\bf E$ corresponds to trivial topologies $k=0$, while generally \cite{David2},
\be
\label{se} F^a_{i0}\equiv E_i^a=\dot N(t) ~(D_i (\Phi_k^{(0)})~ \Phi_{(0)})^a= P_N \frac {\alpha_s}{4\pi^2\epsilon} B_i^a (\Phi _{(0)})= (2\pi k +\theta) \frac {\alpha_s}{4\pi^2\epsilon} B_i^a(\Phi_{(0)}).
       \ee
This field is "transverse":
$$  D ~E =0    $$
due to Eq. (\ref {homo}) (the Gauss law constraint) satisfied with YM potentials (zero mode solutions) $Z^a$, (\ref {zero}).

The {\it Appendix 1} is envisaged in the present article. 
\par
It will be devoted to same aspects of the Euclidian instanton theory \cite{Bel}.

In particular, tunnelling between "in" and "out" vacua (referring, respectively, to the $ x\ \to\pm \infty$ in the Euclidian space $E_4$) will be compared with the one in the Minkowskian Higgs model quantized by Dirac \cite{Dir} 
(involving the Pontryagin index $\nu$, (\ref{Poi})).
\section{Nontrivial topological dynamics in  Minkowskian Higgs model quantized by Dirac.}
\subsection{Collective vacuum modes arise at Gauss-shell reduction of Minkowskian Higgs model with BPS monopoles.}
In Introduction we have already discussed the outlines appearing collective vacuum modes $Z^a$, (\ref{zero}), in  the Minkowskian Higgs model with vacuum BPS monopole solutions  quantized by Dirac \cite{Dir}. 

\medskip
In details, this looks as following.

In  the Minkowskian Higgs model with vacuum BPS monopole solutions (including topologically trivial YM BPS monopole background $\Phi ^{(0)}$), the YM Gauss law constraint (\ref {Gauss}) may be satisfied with the covariant (Coulomb) gauge \cite{LP2,LP1}
\be
\label{Aparallel}
A^{a\parallel}\sim [D^{ac}_i(\Phi ^{(0)})A_c^{i ~(0)}]=0\vert _{t=0}.
\ee
In particular, in the enumerated model, the YM Gauss law constraint (\ref {Gauss}) may be resolved in terms of topological Dirac variables $\hat A^D_i$ ($i=1,2$) \cite{LP1} \footnote{According \cite{rem2,Pervush3},
$$ \hat A_\mu = g \frac {A_\mu^a\tau _a}{2i \hbar c}. $$
}
\bea
\label{degeneration}
\hat A_i^D = v^{(n)}({\bf x}) T \exp \left\{\int  \limits_{t_0}^t d {\bar t}\hat A _0(\bar t,  {\bf x})\right\} \left({\hat A}_i^{(0)}+\partial_i\right ) \left[v^{(n)}({\bf x}) T \exp \left\{\int  \limits_{t_0}^t d {\bar t} \hat A _0(\bar t,{\bf x})\right\}\right]^{-1}, \eea 
with  the symbol $T$  standing for   time ordering  the matrices under the exponent sign. \par
As it was discusses in Ref. \cite{rem2} (repeating the arguments \cite{David2,Azimov}), the topological Dirac variables (\ref{degeneration}), involving Gribov topological Gribov multipliers $ v^{(n)}({\bf x})$, are gauge invariant and transverse functionals of YM fields $A$.\par 
Herewith for topologically trivial multipliers $ v^{(0)}({\bf x})$, functionals (\ref{degeneration}) satisfy the gauge (\ref{Aparallel}), while at $n\neq 0$ they satisfy the Coulomb gauge \cite{Pervush2}
\be
\label{transv}
D_i^{ab} (\Phi _k^{(n)}){A}^{i(n)}_b =0 
\ee 
(implicating topologically nontrivial YM BPS monopole modes $\Phi _k^{(n)}$).
\par
\medskip
Upon fixing the Coulomb gauge (\ref{Aparallel}), (\ref{transv}) for topological Dirac variables (\ref{degeneration}), the YM Gauss law constraint (\ref {Gauss}) turns into the homogeneous second-order differential equation
\be
 \label{odnorod}
 (D^2)^{ab} \Phi_{b(0)}=0,
\ee 
permitting $Z^a$ as its solutions \footnote{The just described method resolving the YM Gauss law constraint was proposed at the first time by Polubarinov \cite{Polubarinov}. It consists, generally, in straight solving constraint equations: classical  equations on field components having equal to zero canonical momenta. }. \par
\medskip
Now let us return to Eq. (\ref{rot}), described collective solid rotations inside the Minkowskian BPS monopole vacuum suffered the Dirac fundamental quantization \cite{Dir}.

To ground  that it is correct, let us appeal to  classical solid mechanics.  \par
  It is well known (see e.g. \S 32 in \cite{Landau1}) that the general solid Lagrangian has the look 
\be \label{tv.telo}
L_s= \frac{\mu V^2}{2}+ \frac{1}{2} I_{ik} ~\Omega_i \Omega_k -U
\ee  
in a chosen rest reference frame. \par  
Herewith $\mu$ is the mass of the given solid:
$$\mu =\sum_a m_a,$$
with $m_a$ being the mass of the $a$$^{\rm th}$ "particle";   $V$ is the  velocity of the c.m. translational motion; $\vec \Omega$  is the angular  velocity of its rotary motion; $U$  is the potential energy of the solid. \par
In conclusion, the tensor $ I_{ik} $ is the \it tensor of inertia momenta \rm (or simply, \it inertia tensor\rm):
\be
 \label{tv.telo1}
 I_{ik} = \sum_a m_a (x_{al}^2 \delta_{ik}- x_{ia} x_{ka}).
\ee
\medskip We can compare both Eqs.: (\ref{rot}) and   (\ref{tv.telo}) \cite{Landau1}.

This comparison shows  that the role of the inertia tensor $ I_{ik} $ is played, in the Minkowskian Higgs model quantized by Dirac and involving vacuum BPS monopole solutions, by the value $I$ \cite{David2}, (\ref{I}), of the vacuum rotary momentum. 
Simultaneously,  the role of the angular  velocity $\vec \Omega$ is played therein by ${\dot N} (t)$. \par  
\medskip

The way deriving Eq. (\ref{I}) for the vacuum "rotary momentum" $I$ was outlined in Ref. \cite{David2}: 
\be
  \label{kt}
I\simeq \frac{4\pi^2}{\alpha_s}\int\limits_{\epsilon}^{R}dr
~\frac{d}{dr}(r^2\frac{d}{dr}f_0^{BPS}(r)); \quad R\to \infty. 
\ee
This Eq. implicates the "Higgs" BPS monopole {\it ansatz} \cite{LP2,LP1}
$$ f_0^{BPS}(r)=\left[  \frac{1}{\epsilon\tanh(r/\epsilon)}-\frac{1}{r}\right].  $$
Herewith the typical size $\epsilon$ of ("Higgs", "YM") BPS monopoles may be set via \cite{LP2,LP1}
\be 
\label{lim}
\frac{1}{\epsilon}\equiv\frac{gm}{\sqrt{\lambda}},
\ee
with $m$ and $\lambda $ being, respectively, the Higgs mass and selfinteraction constant. \par  
As it can be read   from (\ref{I}), $\epsilon \sim V^{-1}$.\par  
More exactly, this dependence may be given  \cite{LP2,LP1, David2} as
\be 
\label{masa}  \frac{1}{\epsilon}=\frac{gm}{\sqrt{\lambda}}\sim \frac{g^2<B^2>V}{4\pi}. \ee
Latter Eq. comes from evaluating \cite{David2} the vacuum "magnetic" energy referring to the BPS monopole configuration:
\be 
\label{magn.e1}
\frac{1}{2}\int \limits_{\epsilon}^{\infty } d^3x [B_i ^a(\Phi_k)]^2 \equiv \frac{1}{2}V <B^2> =\frac 1{2\alpha_s}\int\limits_{\epsilon}^{\infty}\frac {dr}{r^2}\sim \frac 1 2  \frac 1
{\alpha_s\epsilon}= 2\pi \frac{gm}{g^2\sqrt{\lambda}}=\frac {2\pi} {g^2\epsilon} 
\ee  
(at setting $c=\hbar=1$ in the definition of $\alpha_s$).\par  
\medskip
Draw our attention also to the $<B^2>^{-1}$ dependence of the rotary momentum $I$, (\ref{I}).\par 
This is the specific trace of manifest superfluid properties \cite{rem1, rem2} of the Minkowskian Higgs model quantized by Dirac \cite{Dir}.

More precisely, this dependence is stipulated by the {\it Bogomol'nyi  equation } \cite{rem1, LP2,LP1} 
\be
\label{Bog}
 {\bf B} =\pm D \Phi: 
\ee
implicating the Higgs isomultiplet $\Phi$ (having the shape of vacuum BPS monopole solutions), characterizing \cite{rem1} the superfluid properties of the Minkowskian BPS monopole vacuum.

\medskip
In the infinite spatial volume limit $V\to\infty$, the rotary momentum $I$, (\ref{I}), disappears with disappearing the BPS  monopoles size $\epsilon$ via (\ref {masa}).

This implies also suppressing, in the infinite spatial volume limit, the action functional (\ref{rot}), describing collective solid rotations  of the Minkowskian BPS monopole vacuum suffered the Dirac fundamental quantization \cite{Dir}.

\medskip
Due to the variation principle, the action functional (\ref{rot}) yields, as it is easy to see,  the YM Gauss law constraint (\ref{homo}).

This confirms the rightfulness of the Gauss-shell reduction of the Minkowskian Higgs model.

\medskip
Generally speaking, this Gauss-shell reduction of the Minkowskian Higgs model, involving collective zero mode solutions $Z^a$, (\ref{zero}), is one of the  features of the Dirac fundamental quantization \cite{Dir} distinguishing it from the "heuristic" Faddev-Popov (FP) quantization \cite{FP1}, implicating FP path integrals with gauge fixing.

As it was demonstrated in the recent paper \cite{fund} (repeating the arguments \cite{Arsen}), the presence of collective (vacuum) excitations in a gauge theory (collective zero mode solutions $Z^a$, (\ref{zero}), are the example of such collective excitations) involves violating the {\it gauge equivalence theorem}  \cite{Taylor, Slavnov1} between the Dirac \cite{Dir} and FP \cite{FP1} quantization approaches.

The both methods prove to be of equal worth when S-matrixes for quantum fields on-shell are in question.

But  when composite fields are present in a gauge model (for instance, the collective excitations or bound states), the gauge equivalence theorem \cite{Taylor, Slavnov1} is violated \cite{fund, Arsen} and various {\it spurious} Feynman diagrams  (SD) \cite{Pervush2, Nguen2} appear on the level of appropriate Green functions:
$$  ({\rm FR})^F+ ({\rm SD})\equiv ({\rm FR})^* \quad ({\rm for~ S- matrices~ with ~composite~ fields});  $$
$$({\rm FR})^F+ ({\rm SD})\equiv ({\rm FR})^* \quad ({\rm for~ Green~ functions)},
$$
where the Feynman rules $({\rm FR})^F $ refer to the FP gauge fixing method \cite{FP1}, while the Feynman rules $({\rm FR})^* $ refer to the Dirac fundamental quantization approach \cite{Dir}.
The latter rules implicate (topological) Dirac variables.

As it was demonstrated in Refs. \cite{Pervush2, fund,  Nguen2}, spurious Feynman diagrams  (SD) take account of the manifest relativistic covariance of  (topological) Dirac variables \cite{Pervush2, fund, Polubarinov, Nguen2} (we refer our readers to Ref. \cite{fund} for details).
\subsection{Properties of topological variable $N(t)$.}
The topic of this subsection will be investigating properties of the topological variable $N(t)$ (with its time derivative $\dot N(t)$, entering the expression (\ref{rot}) for the BPS monopole vacuum rotary action functional).

Herewith $N(t)$ plays the crucial role in the (topologically nontrivial) dynamics inherent in the Minkowskian Higgs model with vacuum BPS monopoles quantized by Dirac \cite{Dir}. Therefore the present subsection is most important.

\medskip
As it was demonstrated in Refs.  \cite{LP2,LP1,Pervush2}, the topological variable $N(t)$ may be specified via the relation
\bea
\label{winding num.}
\nu[A_0,\Phi^{(0)}]&=&\frac{g^2}{16\pi^2}\int\limits_{t_{\rm in} }^{t_{\rm out} }
dt  \int d^3x F^a_{\mu\nu} \widetilde{F}^{a\mu \nu}=\frac{\alpha_s}{4\pi}  \int d^3x F^a_{i0}B_i^a(\Phi^{(0)})[N(t_{\rm out}) -N(t_{\rm in})]\nonumber \\
 &&  =N(t_{\rm out}) -N(t_{\rm in})= \int\limits_{t_{\rm in} }^{t_{\rm out} } dt \dot N(t), 
 \eea 
taking account of the natural duality between the  tensors $ F^a_{i0}$ and $ F^a_{ij}$ and the normalization \cite{Pervush2} $$
 \frac{g^2}{16 \pi^2}\int d^3 x D_i^{ab}({\Phi})\Phi^b_0 B_i^a({\Phi})=1; \quad  \alpha_s\equiv g^2/4\pi.$$ 

Herewith $\nu[A_0,\Phi^{(0)}]$ is referred to as the {\it vacuum Chern-Simons functional}, implicating the asymptotical states  "{in}"  and "{out}" taking in the time instants $t_{\rm in}$ and  $t_{\rm out}$, respectively. 
Herewith the mentioned time instants may be chosen to be \cite{Pervush3} $t_{\rm in}=T/2=- t_{\rm out}$ (with an arbitrary $T$).

In Ref. \cite{Pervush2}, the vacuum Chern-Simons functional (Pontryagin index) $\nu [A]$ was defined as
\be \label{Ch-S} \nu[A]=\frac {g^2}{16\pi ^2}\int\limits _{t_{\rm in}} ^{t_{\rm out}} dt\int \limits _ {V} d^3 x F_{\mu \nu}^a 
{\tilde F}^{\mu \nu}_a = X[A_{\rm out}]-X[A_{\rm in}]= n(t_{\rm out})-n(t_{\rm in});\quad n_{\rm in,~out}\in {\bf Z};\ee 
with 
\be \label{wind} X[A]=-\frac {1}{8\pi ^2}\int\limits _ {V} d^3 x \epsilon ^{ijk} {\rm tr}~ [{\hat A}_i \partial_j{\hat A}_k-  \frac {2}{3}{\hat A}_i{\hat A}_j {\hat A}_k],\quad A_{\rm in,~out}= A(t_{\rm in,~out},{\bf x}), \ee 
being the \it topological winding number functional of  gauge  fields\rm. 

It follows from (\ref{winding num.}), (\ref{Ch-S}) and (\ref{wind}) that $N(t)$
may be treated as the noninteger degree of the map referring to the $U(1)\to SU(2)$ embedding \footnote{The Pontryagin degree of a map theory is stated enough good in the monograph \cite{Al.S.} (in \S T21), and we recommend this monograph to our readers for studying the question.

Now note only  that degrees of maps take integers according the Pontryagin theory.}.

\medskip
As we have  discussed above, the topological dynamical variable $N(t)$ may be represented in the shape (\ref{tbo}) \cite{Pervush3}.

In this case, as it is easy to see \cite{Pervush3},  $N(t)$  takes integers  as $t_{\rm in}=T/2=- t_{\rm out}$:
\be
\label{tbc}
N(\pm T/2)=n_{\pm}; \quad n_{\pm}\in {\bf Z}. \ee
Latter Eq. may be considered as an original "boundary condition" imposed on $N(t)$. 

As a consequence of (\ref{tbo}), one comes to Eq. (\ref {Poi}) for $\dot N(t)$. 

\bigskip
Indeed, and this has a crucial importance, from the QFT point of view, it should be set $T\to \infty$.

This setting has profound roots in QFT associated with the  {\it Haag theorem} \cite{Logunov}.

It turns out that at considering asymptotical states for quantum fields in finite time instants $t_{\rm in,~out}$ one encounters lot of problems and troubles.    

The thing is that \cite{Logunov} the interaction picture for quantum fields is based on assuming that, in QFT, generalized coordinates (quantum fields) interacted $\phi$ and appropriate generalized momenta $\pi$ (canonically conjugate to $\phi$) are related with an unitary transformation with "free" variables $\phi_0 $ and $\pi_0$ in the fixed time instants $t$. 

For instance,
\be \label{unitas}  \phi (t,{\bf x}) = V(t)\phi_0 (t,{\bf x}) V^{-1}(t)      \ee
for generalized coordinates.

In  this situation, as the Haag theorem shows \footnote{Briefly, in the relativistic case, the Haag (-Hall) theorem may be formulated as follows \cite{Logunov}.

One assumes that, in the fixed time instants $t$, the fields $\phi (t,{\bf x})$, $\dot \phi (t,{\bf x}) $ and  $\phi_0 (t,{\bf x})$, $\dot \phi_0 (t,{\bf x})$ (the existence of such fields is postulated) form two irreducible systems of quantum fields in the Hilbert spaces: respectively, $\cal H$ and ${\cal H}_{(0)}$, of  quantum states.

Then, in addition to (\ref{unitas}), it is assumed
$$ \dot \phi (t,{\bf x}) = V(t)\dot \phi_0 (t,{\bf x}) V^{-1}(t).           $$
In  this case four first Wightman functions of the fields $\phi (t,{\bf x})$ and $\phi_0 (t,{\bf x})$ coincide in the both theories.

Moreover, if $\phi_0 (t,{\bf x})$ is a free field with  the mass $m\geq 0$,  then $\phi (t,{\bf x})$ is also the free field with  the mass $m$ and the both theories coincide to within an unitary  transformation $V$.}, the quested QFT becomes trivial if the claim for quantum fields to be relativistic invariant is added to the standard claims to these fields in QFT \cite{Logunov} (the so-called {\it Wightman axioms}). In other words, quantum fields prove to be free.

More exactly \cite{Logunov}, in  this case (at $t\neq \pm \infty$), there are no good specified matrix $V(t)$ for the abovementioned unitary transformation over the Hilbert space $\cal H$ of  quantum states.

To ground the latter statement, let us analyse briefly the canonical Lagrangian formalism in QFT.

The  canonical quantization scheme is based \cite{Landau1} on the classical Lagrangian 
$$   {\cal L}= {\cal L}(\phi_\alpha (x), \frac { \partial \phi_\alpha (x)}
{\partial x^\nu}),    $$
which is a function of (quantum) fields $\phi_\alpha (x)$ and their first derivatives.

The canonically conjugate "momentum" to the field $\phi_\alpha (x)$ in the time instant $t$ is given by Eq.
$$   \pi_\alpha (t, {\bf x})= \frac {\partial {\cal L}}{\partial \frac {\partial \phi_\alpha (t, {\bf x})} {\partial t}}.    $$ 
It is postulated that canonically conjugate fields $\phi_\alpha $ and $\pi_\alpha $ are the elements of an algebra determined by canonical commutation relations (CCR) in the case of bosonic  fields \footnote{ They would be replaced with anticommutation relations in the case of fermionic fields.} 
$$
[ \phi_\alpha (t, {\bf x}),\phi_\beta (t, {\bf y})]=[ \pi_\alpha (t, {\bf x}), \pi_\beta (t, {\bf y})]=0;
$$
\be\label{CCR}
[\phi_\alpha (t, {\bf x}),\pi_\beta (t, {\bf y})]= i\delta_{\alpha \beta } 
\delta ({\bf x}-{\bf y}).
\ee
As a result, QFT may be formulated as quantum mechanics of a system with  the infinite number of degrees of freedom (herewith $\bf x$ and $\bf y$ plays the role of the "number" of the given generalized coordinate and momentum).

It is convenient also to deal  with a countable basis instead of a continuous.

For this purpose, in   \cite{Logunov} it was proposed to introduce an orthonormalized system of functions $ h_\nu({\bf x})$ in the three-dimensional space (for instance, there can be Hermit functions $N_\nu e^{-x^2/2} H_{\nu_1\nu_2\nu_3}({\bf x})$)\footnote{Indeed, $H_{\nu_1\nu_2\nu_3}({\bf x})= H_{\nu_1}({\bf x})H_{\nu_2}({\bf x}) H_{\nu_3}({\bf x})$, where $H_{\nu_1\nu_2\nu_3}({\bf x})$ ($i=1,2,3$) are the three-dimensional Hermit polynomials.}  and to specify "coordinates" and "momenta" with Eqs.
$$ Q_{\bf n} (t)=\int  \phi_\alpha (t, {\bf x}) \bar  h_\nu({\bf x}) d^3x, $$
\be\label{Pt}
P_{\bf n} (t)=\int  \pi_\alpha (t, {\bf x})   h_\nu({\bf x}) d^3x,
\ee
with $\bf n$ being the composite discrete index: ${\bf n}\equiv(\alpha,\nu)$.

It may be shown herewith \cite{Logunov} that CCR (\ref{CCR}) take the look 
$$
[Q_{\bf n} (t), Q_{{\bf n}'} (t)]= [P_{\bf n} (t), P_{{\bf n}'} (t)]=0;
$$
\be\label{CCR1}
[Q_{\bf n} (t), P_{{\bf n}'} (t)]=i\delta_{{\bf n}{\bf n}'}.
 \ee
Further, the abstract algebra of elements $ Q_{\bf n}$, $ P_{\bf n}$ is realized satisfying (\ref{CCR1}).

It is, indeed, an algebra of unrestricted operators in the Hilbert space $\cal H$ of state vectors.

In the case of a system with finite degrees of freedom, some two irreducible representations of CCR (\ref{CCR1}) realized via selfconjugate (Hermitian) operators in the Hilbert space $\cal H$ are unitary equivalent \footnote{In the latter statement is the substance of the so-called {\it von Neumann theorem} (see Theorem 6.14 in \cite{Logunov}).

Let us an irreducible {\it Weyl system} be given. It is set over the Hilbert space $\cal H$ and involves CCR represented in the exponential form for a quantum  system with some $n$ degrees of freedom.

Any Weyl system always comes to the pair of Abelian $n$-parameter groups $U(a)$ and $V(b)$: 
$$  U(a)= e^{iap}; \quad    V(b)= e^{ibq},     $$
respectively.

Here $a,b\in {\bf R}^n$, and
$$  ap\equiv \sum\limits_{j=1}^n a_j p_j; \quad bp\equiv \sum\limits_{j=1}^n b_j q_j     $$
for a system of (generalized) coordinates $q \equiv (q_1,\dots q_n)$ and (generalized) momenta $p \equiv (p_1,\dots p_n)$ conjugate to former.

It is well known \cite{Landau3} that the operators $p$ and $q$ are selfconjugate (Hermitian).

Herewith also 
$$  U(a) U(a')=U(a+a'); \quad V(b) V(b')= V(b+b'),                  $$
and the commutation relations between $p$ and $q$ come to Eq. \cite{Logunov}
$$  U(a) V(b) =e^{iab} V(b) U(a).     $$
\medskip
The claim for a Weyl system $\{ U(a), V(b)  \}$ to be irreducible comes to the claim that there are no nontrivial closed subspace in the Hilbert space $\cal H$ invariant with respect the operators $ U(a), V(b)$.

\medskip
On the other hand, if 
$$  p_i= -i \frac \partial {\partial q_i}$$
for any i, and if the vectors $p$ and $q$ are defined over the  Lebesgue space ${\cal G}^2({\bf R}^n)$, the  operators $p$ and $q$ form the representation for CCR referred to as the {\it Schr${\rm \ddot o}$dinger representation} \cite{Logunov}.

\medskip
In the above terms, the von Neumann theorem may be formulated as follows \cite{Logunov}.

Any  irreducible Weyl system with $n$ degrees of freedom is unitary equivalent to the Schr${\rm \ddot o}$dinger representation in the Lebesgue space ${\cal G}^2({\bf R}^n)$. Any  reducible Weyl system (with $n$ degrees of freedom) is the direct sum of irreducible representations and thus it is a multiple of the Schr${\rm \ddot o}$dinger representation. }. 

In particular, an unitary operator $V(t_2,t_1)$ exists associated the operators $P_{\bf n}$ and $Q_{\bf n}$ referring to different time instants:
$$ Q_{\bf n}(t_2)= V(t_2,t_1) Q_{\bf n}(t_1) V^{-1}(t_2,t_1),  $$ 
\be
\label{V}
P_{\bf n}(t_2)= V(t_2,t_1) P_{\bf n}(t_1) V^{-1}(t_2,t_1).
\ee 
But the said is not correct for systems  with infinite numbers of degrees of freedom.

Moreover, linear canonical transformations (i.e. transformations of variables $P_{\bf n}$ and $Q_{\bf n}$ maintaining the look of CCR (\ref{CCR1})) don't correspond, generally speaking, to an  unitary equivalence transformation \footnote{To understand this idea, we recommend our readers to study the arguments given in Exercise 7.24 in Ref. \cite{Logunov}.}.

\medskip
Among the infinite set of non-equivalent representations of CCR, one can pick out the {\it Fock representation} (see \S8.4 in \cite{Logunov}) in the theory of {\it free} fields.

This picking out may be achieved with the aid of the additional claim that there is the unique normalized relativistic invariant state (vacuum) $\Psi_0$ annihilated under acting the positive frequencies operators:
\be \label{Fock}
\tilde \phi^{(+)} (p)\Psi_0=0.
\ee
This claim is equivalent to existing the vacuum $\Psi_0$.

The Fock representation is specified in the unique way (to within the unitary equivalence).

On the face of it, it can be assumed (and it is assumed often) that one can pick out the { Fock representation} also in the theory of {\it interacting} fields. Herewith that representation would involve a physical vacuum maintained in the time.

\bigskip Indeed, the Haag theorem shows that it is not correct and that one would utilize another representations than the  Fock ones for CCR (herewith there is the "actual" infinite tower of "bare" particles in each state in such representations). 

What does this theorem say? Let us turn to the presentation of the Haag theorem (in the relativistic theory!) in the monograph \cite{Logunov}  (see Theorem 9.28 p. 352 therein). So, same two scalar neutral Wightman fields $\phi(x)$  and $\phi_{(0)}(x)$ are considered which act into the appropriate Hilbert spaces $\cal H$ and ${\cal H}_{(0)}$.  Suppose further that in any time instant $x_0=t$ the fields $\phi(t,{\bf x})$ and $\dot \phi(t,{\bf x})$, and also $\phi_{(0)}(t,{\bf x})$ and $\dot \phi_{(0)}(t,{\bf x})$  (the existence of such  fields is postulated!) form two irreducible systems of fields in  $\cal H$ and $ {\cal H}_{(0)}$, respectively. Finally
, let us suppose that in the time instant $x_0=t$ these two systems are related by the (time dependent) unitary transformation $V\equiv V(t)$ 

\be \label {unit1} \phi(t,{\bf x})= V(t)  \phi_{(0)}(t,{\bf x})V(t)^{-1}, \quad  \dot\phi(t,{\bf x})= V(t)  \dot \phi_{(0)}(t,{\bf x})V(t)^{-1}
\ee
In this case the first four  of the fields $\phi(x)$  and $\phi_{(0)}(x)$ coincide in the both theories. If, besides that, 
$\phi_{(0)}(x)$ is a free field with the mass $m\ge 0$, then  the field $\phi(x)$ {\it is also the free field with the mass} $m$ and the both theories {\it coincide} \footnote{The free scalar field $\phi_{(0)}$ satisfies the Klein–Gordon equation 

$$ (\square +m^2)\phi_{(0)} =0$$ and possesses the Lorentz invariant Lagrangian,  the Lorentz invariant Wightman functions, the Lorentz invariant spectrum while it is { \it Poincare-covariant}:

$$ U(a,\Lambda)\phi_{(0)}(x) U^{-1}(a,\Lambda)=\phi_{(0)}(\Lambda x+a),$$ where $U(a,\Lambda)$  is the unitary representation of the  Poincare group.

 Thus all the $n$-point Wightman functions of a free field, $W_n (x_1,...,x_n)$, are Lorentz invariant distributions. 

Moreover,  a free field  $\phi_{(0)}(x)$ satisfies the  Wick's theorem:
$$ W_n (x_1,...,x_n) =\sum_{\rm all~pairings} \prod W_2(x_{ik}, x_{jk})$$ This means that $W_2$ fix the mass and spectrum, while everyone else $W_n$ are expressed through $W_2$. Herewith there are no additional independent Lorentz invariants for $n \ge 5$!

\medskip  However when $n\ge 5$ and when the field  $\phi_{(0)}(x)$ is not free,  lot of problems arise. Briefly speaking, for $n\ge 5$, Lorentz transforms cannot bring an arbitrary configuration of points to the equal times.
}.   

\medskip Returning, in the framework of the Haag theorem, to the problems with the Fock representation, we now can show how the above mentioned tower of ”bare” particles arises. To begin with, note that any interaction is absent in the phrasing of the Haag theorem in \cite{Logunov} if the field  $\phi_{(0)}(x)$ is free. In this case, the unitary operator $V(t)$  has nothing to do with the interaction!

Now we suppose that interactions are exist after all in the model with the scalar neutral field. Then the physical vacuum (maintained in the time) generates a state $| \Psi_{\rm phys}>$. Its expansion by the Fock basis of the  free field  $\phi_{(0)}(x)$ has the look
\be \label{Fock b} 
| \Psi_{\rm phys}>= \sum _{n=0}^\infty \int d^4 x_1...  d^4 x_n c_n (x_1,... x_n) \phi_{(0)}(x_1)... \phi_{(0)}(x_n).
 \ee It is remarkably that the coefficients $c_n$ are now the functions of many variables. Eq. (\ref{Fock b}) is just the exact form of the statement that any physical state (in the scalar field model) contains the tower of free (bare) modes.

\medskip   The said can be illustrated with the aid of the spectral decomposition of the two-point Wightman function for the scalar field $\phi (x)$. Namely,  
$$ <\Psi_0, \phi (x) \phi (y), \Psi_0>. $$ This Lorentz-covariant functions permits the spectral representation 

\be\label{spectr} W_2(x-y)=\int _0^\infty d \mu_2 \rho(\mu_2) \Delta^+(x-y,\mu_2),\ee with $\rho(\mu_2)\ge 0$ being the spectral density and  $\Delta^+$ being the "free" two-point function of the mass $\mu_2$.

If a one-particle state of the mass $m$ is present in the spectrum, then 
\be \label{romu} \rho(\mu_2)=Z\delta( \mu_2-m)+\rho_{\rm cont}(\mu_2), \quad \rho_{\rm cont}\ge 0.\ee  Then 
\be \label{Zfactor} W_2(x-y)= Z \Delta^+ (x-y, m)+ \int _{2m}^\infty d \mu_2 \rho_{\rm cont}(\mu_2) \Delta ^+(x-y,\mu_2)\ee

\medskip  Now consider the role of the factor $Z$ in the decomposition (\ref{Zfactor}). Indeed, $Z$ is the renormalization coefficient for the bare scalar field $\phi_{(0)}$ to the dressed field $\phi$ ($Z_\phi$ in the notations \cite{Cheng}):

\be \label{rfi} \phi =Z^{-1/2} \phi_{(0)}. \ee The authors of \cite{Cheng} give an enough simple formula for $Z$. Let $\Sigma '(\mu^2)$ being the logarithmically divergent contribution into the self-energy of the scalar field $\phi$. Then 

$$ Z=[1-\Sigma '(\mu^2)]^{-1}= 1+\Sigma '(\mu^2)+ O(\lambda _0^2),$$ with $\lambda _0$ being the coupling constant. 
 
Thus Eq. (\ref{Zfactor}) states that $\phi$ creates the one-particle state with the amplitude $Z$ and, simultaneously, multi-particle continuous excitations. On the contrary, in the free field $\phi_0$ model, $Z=1$, $\rho_{\rm cont}(\mu_2)=0$.  

\medskip Let us specify now the renormalized field $\phi$ as in (\ref{rfi}) \cite{Cheng}. Then

$$<\Psi_0|\phi(x) \phi(y)|\Psi_0>= \Delta ^+(x-y,m)+{\rm continium}.$$  In other words, the two-point Wightman function contains the "free part" plus the continuous "tail".

\medskip Now we can discuss directly where the "unFockian" nature of the representation $W_2(x-y)$, (\ref{spectr}), manifests itself. If this representation were Fockian, the model Hamiltonian would decompose into the sum of oscillators; the spectrum would consist of integer positive multiples of the mass $m$; the  spectral density would be merely the sum of $\delta$-functions

Once the interaction is turned on, the $\rho_{\rm cont}(\mu_2)$ item appears in (\ref{romu}); the multi-point  functions now do not satisfy the Wick's theorem; the space of states is not unitary equivalent to the  Fock space.

\medskip Note additionally that Lorenz-Poincare properties of the  dressed field $\phi$ and its ( two-point) Wightman functions are maintained at going over from the free field $\phi_0$. In particular,  $\phi$ is transformed as a scalar.

The mentioned difficulties associated with the Haag theorem may be avoid at utilizing the asymptotic interaction picture in which interacting fields are equated in the time instants $t=\pm \infty$ instead their equating in a finite time instants $t$.  

In this case,  interacting, $\phi(x)$, and free, $\phi^{\rm in}(x)$, fields are related as \cite{Logunov}
\be \label{fin}
\phi(x)=S^* T(S \phi^{\rm in}(x)), 
\ee 
with $T$ standing for time ordering and $S$ being the appropriate scattering matrix \footnote{The properties of S-matrices were described good in the monographs \cite{Logunov, Bogolubov-Shirkov}.

In particular, it is worth to note that S-matrices are always Poincare invariant unitary operators in the Fock  space $F$ of a system of free relativistic particles (see e.g. \S7.3 in \cite{Logunov}): for instance, this space may be identified with the space of fallen particles.

Herewith the irreducible system of (Wigthman) free ("in") fields $ \phi^{{\rm in (\chi)}}$ with CCR 
$$  [\phi^{{\rm in(\chi)}}_l(x), \phi^{{\rm in}(\chi')}_{l'}(y)] = \frac 1 i D_{ll'}^{(\chi \chi ')} (x-y)        $$
(where in the denotations \cite{Logunov}, $l$, $l'$ being Lorenz indices and $\chi$, $\chi'$ being indices specifying types of fields; $D(x)$ is the appropriate Pauli-Jordan permutation function)
at the normal spin-statistic connection (referring to the time instant $t=-\infty$) acts in  $F$.

As it is well known, the S-matrix permits its (functional) expansion \cite{Efimov}
$$  S= \sum\limits_ {n=0}^\infty  \frac 1 {n!} \int dx_1  \dots \int dx_n \sum\limits_ {\alpha_1\dots \alpha_n} S_ {\alpha_1\dots \alpha_n} ^{(n)}(x_1, \dots  x_n)\times$$
$$ :\phi_{\alpha_1}(x_1) \dots \phi_{\alpha_n}(x_n), $$
involving composite indices $\alpha_i$ ($i=1, \dots n$) consisting of $l$, 
$\chi$ and "in" ("out") indices \cite{Logunov}.

Herewith the coefficient functions $ S_ {\alpha_1\dots \alpha_n} ^{(n)}(x_1, \dots  x_n)$ in this expansion (due to the Wick rules 
\cite{Bogolubov-Shirkov},  normal (N) ordering of quantum asymptotical fields $\phi_{\alpha_i}(x_i) $ would be taking into account as well as various connections between these fields) are integrable functions belongs to the classes of distributions (as a rule, there are  Schwarz distributions over test functions with  compact definition regions or  distributions over test functions with  moderate increasing: see Chapter 2.1 in \cite{Logunov}).

The said just ensures good specifying asymptotical states in (\ref{fin}).
}.

Latter Eq. is equivalent to the {\it Yang-Feldman equation} \cite{Logunov, Ryder} 
\be \label{Yang-Feldman}
\phi^{(\chi)}_{l}(x) =\phi^{{\rm in}(\chi)}_{l}(x)+ \sum\limits_{\chi' l'} \int D_{ll'}^{{\rm ret}(\chi \chi ')} (x-y) J^{(\chi') l'} (y) dy.       
\ee 
This Eq. implicates the {\it retarded} Green function \cite{Logunov, Bogolubov-Shirkov}
\be \label{ret}
D^{\rm ret} (x)= \int \frac 1 {m^2-p^2+i0\epsilon (p^0)} e^{ipx}d_4p, 
\ee
where \cite{Bogolubov-Shirkov}
$$ \epsilon (\alpha)= \frac 1 {\pi i} {\cal R} \int \limits_{-\infty}^{\infty }e^{i\alpha\tau} \frac{d\tau }\tau =\left\{\begin{array} {l@{}r@{\ \le t \le\ }l}
 1~~ {\rm at}~~ \alpha >0\\ -1 ~~ {\rm at}~~ \alpha <0  \end{array} \right\}.
$$
(with ${\cal R} $ denoting the main  value).

For the field $\phi^{{\rm out}(\chi)}_{l}(x)$ (at $t=+\infty$), one would replace $ D^{\rm ret} $ with the {\it advanced} Green function 
\be \label{adv}
D^{\rm adv} (x)= \int \frac 1 {m^2-p^2-i0\epsilon (p^0)} e^{ipx}d_4p
\ee
in the Yang-Feldman equation (\ref{Yang-Feldman}).

Note also that, from the viewpoint of (relativistic) quantum mechanics, attributing free fields to the time instants $t=\pm \infty$ corresponds just to the interaction picture in such its picture \cite{Bogolubov-Shirkov} when an interaction $H_{\rm I}$ is "switched on" adiabatically at  $t=- \infty$  and is "switched off" also adiabatically at  $t=+ \infty$.

Herewith, at denoting as $ \Phi(- \infty)$ the amplitude of the initial ("in") state and as  $ \Phi(+ \infty)$ the amplitude of the final ("out") state (in the appropriate Fock space $F$), takes place the relation
\be \label{predst. vsaimod.} 
\Phi(+ \infty)= S\Phi(- \infty)
\ee
between these amplitudes, implicating the scattering matrix $S$.

The interdependence between the relations (\ref {predst. vsaimod.}), (\ref {fin})
and the Yang-Feldman equations (\ref{Yang-Feldman}) is highly transparent.
 
\medskip
The just describing procedure \cite{Logunov} referring free fields $\phi^{\rm in}(x)$ and $\phi^{\rm out}(x)$ to the time instants $t=\pm \infty$, respectively (involving the Yang-Feldman equations (\ref{Yang-Feldman})), results (as it may be demonstrated) deleting infrared divergences in  quested QFT (although ultraviolet divergences can occur).

The mentioned  infrared divergences are just associated with the Haag theorem.
The origin of them is in the infinite spatial volume utilized at the proof of this theorem (for instance, in Eqs. (\ref {Pt}) specifying the operators $Q_{\bf n}(t)$ and $P_{\bf n}(t)$; this just causes the bad definition for the operator $V(t_2,t_1)$ because of (\ref {V})).

These divergences can be tamed at a compactification performed, but quested QFT may still be plagued by divergences of other types; for example, irreducible representations for a free scalar field and a $\phi^4$  field in the spatially
compactified (2+1)-dim Minkowski space-time are known to belong to in-
equivalent representations because of ultraviolet divergences that occur at
high energy-momentum \cite{Jaffe} \footnote{The arguments by J. Earman and D. Fraser were  utilized here (visit the site http://philsci-archive.pitt.edu/archive/00002673).}.

\bigskip  We have discussed above the problems and ways  to get around these  problems regarding the Haag theorem in the simple case of the scalar field $\phi$.  In the case of topologically nontrivial field configurations, in particular, YM and Higgs monopole configurations, the situation becomes even more complicated.   Why it is so?

Firstly, the Gauss law constraint \footnote{The Gauss law constraint is the cornerstone of the Dirac fundamental quantization of YMH models with vacuum BPS monopole configuration. The program work in this direction is \cite{Pervush2}. The review of these ideas was given in \cite{fund}. For comparison, the Gauss-shell QED, involving Dirac variables, was constructed in the paper \cite{Pervush3} dating back to 1985 y. The arguments \cite{Pervush3} regarding Gauss-shell QED were repeated not so long ago in the work \cite{abel}.} \be\label{gl} G^a(x) = (D_i E_i)^a(x) - \rho^a(x) = 0 \ee
(involving the "electric" field $\vec E$ and non-Abelian current $\rho$). This is an operator equation which physical states must satisfy:

\be\label{gl1}  G^a(x)|\psi_{\text{phys}}\rangle = 0. \ee  This means that not every state in a formal Fock space is admissible; the physical states form a subspace carved out by an infinite number of local conditions  (at least because the manifest dependence of the Gauss law constraint on the 4-coordinate $x$; thus we deal with the infinite and continuous set of local conditions!). {\it This subspace is by no means Fock space}! Really, a Fock space is built always from the
set of independent harmonic oscillators with a positive norm and independent degrees of freedom in each point of the momenta space. 

 Another situation is in a gauge theory. Due to the Gauss law constraint (\ref{gl}), degrees of freedom cease to be independent. Moreover, the longitudinal and temporal components of the gauge field, which are not physical, should be excluded.  The Dirac quantization scheme developed in the papers \cite{Pervush2,Pervush3} is an important pattern of such removal. As we have shown in Section 1, utilizing the ideas of \cite{Pervush2}, the removal of the longitudinal YM modes from the Gauss law constraint (\ref{Gauss}) results the zero mode solutions (\ref{zero}), generating the nontrivial (vacuum) topological dynamics (the subject of the present study), thread topological defects inside the YM-Higgs vacuum manifold \cite{disc} and the first order phase transition in the studied model ("frozen" in the absolute zero temperature limit $T\to 0$).

\medskip By the way, even at the level of intuition, iz is easy to see that the Fock space results all the possible field configurations, while the Gauss law constraint allows only those configurations which are gauge invariant.  In a lattice approximation it looks as if we have an infinite lattice of oscillators but only those states are allowed {\it where the sum of all oscillators is equal to zero at each point}!

\bigskip Secondly, there is no global decomposition into the modes. More precisely, in the free field theory we have
\be \label{dekompoz} A_\mu(x) = \sum_{\mathbf{k},\lambda} \left( a_{\mathbf{k},\lambda} e^{-ikx} + a^\dagger_{\mathbf{k},\lambda} e^{ikx} \right) \ee  (the decomposition by the momenta $\mathbf k$ and helicities $\lambda$).

But, in a gauge theory, the field $A_\mu(x)$ is not an observable value, its components can be subjected to gauge $SU(2)$ transformations of the 

$$ A '_\mu = v^{-1} (x)
A _\mu v (x)+ v(x) \partial _\mu v^{-1}(x),  \quad v(x)\in SU(2), $$ shape. On the other hand, the physical degrees of freedom are the {\it gauge invariant} combinations. Such are, for example, the Wilson loops. 

Note that  Wilson loops can not be decomposed into the discrete sum of  modes of the $a_{\mathbf{k}}$ shape because of their nonlocal nature, their manifest dependence on  the integration contour. And moreover, they possess no linear structure.  Thus it is impossible to construct the global basis as that  utilized in the decomposition (\ref{dekompoz}) and, as a consequence, it is impossible to construct the Fock space in this case!

\medskip Also, in a gauge (non-Abelian) model, {\it there is no particle number operator}.  This has the look

$$  N = \sum_{\mathbf{k},\lambda} a^\dagger_{\mathbf{k},\lambda} a_{\mathbf{k},\lambda}.   $$ It can exist only if  there exist: the mode expansion, the linear equations of motion, the free fields.

However, in the YM theory, one deal with excitations which are gauge invariant objects: gluonic strings and glueballs.  But these are {\it nonlinear} due to the gluonic selfinteraction. This predetermines the absence of the particle number operator in the Yang-Mills model. 

\medskip   {\it There is no global vacuum in the Yang-Mills theory}. Really, in the free theory,
$$  a_{\mathbf{k},\lambda} |0\rangle = 0.  $$  But in the YM model, the vacuum is the whole family of topological sectors $|n\rangle$ (see, for example, \cite{rem1, Cheng, Ryder} and Appendix in the present study). A "large" (with $n \neq 0$) gauge transformation connects two such sectors.  In particular, in the Euclidian $E_4$ space, the physical vacuum is the $\theta$-superposition \cite{Cheng} 
$$ |\theta\rangle = \sum_{n\in\mathbb Z} e^{in\theta} |n\rangle.   $$  This will be the topic of our Appendix.   $|\theta\rangle$ is not the vacuum vector in the Fock space. 

\medskip  The "small" gauge transformations (with the topological number $n=0$) play also their important role in this context.  These gauge transformations can be tightened to the unit element. They act inside the same topological sector:
$$ \mathcal H_n \xrightarrow{\text{small gauge}} \mathcal H_n.   $$  In this sense, each vector inside the fixed $\mathcal H_n$ can be connected with another one through the suitable (local)   "small" gauge transformations. And this is just the manifestation of the  irreducibility of representation in the sector.

\medskip The important component of (Minkowskian) YM-Higgs theories with vacuum monopole configurations \cite{rem1, Al.S.,Cheng, Ryder}  is the magnetic charge. A detailed theory of magnetic charge was given recently in the paper \cite{rem1} following the arguments  \cite{ Al.S.}. It has the look

\be \label{magn. zarjad} Q_m = \frac{1}{4\pi} \int_{S^2_\infty} \vec{B}\cdot d\vec{S},    \ee where the integration is carried out along the remote sphere $S^2_{\infty}$. The magnetic charge $Q_m$ is a topological invariant. This number depends only on the behavior of the "magnetic" field $\vec{B}$ at the spatial infinity.   Thus it is not a local operator. Indeed, a local operator $O(x)$, which has a compact support,  field on the remote surface $S^2_\infty$ and thus does not change the magnetic charge $Q_m $. Thus
$$  O_{\text{loc}} : \mathcal H_{Q_m} \to \mathcal H_{Q_m}. $$
It is in fact obvious that the sectors with different $Q_m$ are the different irreducible representations. This is so since the topological (magnetic) charge   lies in the center of the observables algebra and its eigenvalues marks  different representations.

\medskip Let us show now that a topological (magnetic) charge, as a topological invariant, is indeed a {\it superselection} operator (in the terminology p.224 \cite{Logunov}). It can be seen that this is a global invariant  if we rewrite it in the form 

\be \label{qum}Q_m \sim \int_{S^2_\infty} \mathrm{Tr}(F\wedge A + \dots).  \ee  Herewith any local observables algebra $\cal U$ commutes with the topological (magnetic) charge $Q_m$. More precisely,  $$ [Q_m,O]=0 $$ for any local operator $O\in  \cal U$. But that's not enough for $Q_m$ to be a superselection operator!
As it is stated in the monograph \cite{Logunov} (p.224),  a superselection operator is such operator which is restricted, belongs to the algebra $\cal B (\cal H)$ of all the linear restricted operators in the arbitrary Hilbert space $\cal H$, multiples of the unit operator in each the coherent space ${\cal H}_\nu$ \footnote{A non-empty set $\cal M$ of non-zero vectors in the Hilbert space $\cal H$ is called {\it interlaced system} of vectors if it cannot be represented as a system of two (or more) non-empty mutually-orthogonal subsets.  In this framework, Lemma 6.3 in \cite{Logunov} states that the Hilbert space $\cal H$ is expanded, in an unique way, into the direct sum of (different from zero) subspaces
$$ {\cal H}=\bigoplus _{\nu \in N}{\cal H}_\nu $$ such that $\cal M$ is the union of subsets ${\cal M}_\nu ={\cal M}\bigcap {\cal H}_\nu$ each of which is total in the appropriate subspace $ {\cal H}_\nu$ and form herewith the interlaced system.

In each ${\cal H}_\nu$  the {\it superposition principle} takes place in a restricted shape (namely, within the subspaces $ {\cal H}_\nu$): a nonzero linear combination of the pure states vectors is again the  pure states vector at the condition that the source vectors lie in the same subspaces $ {\cal H}_\nu$. This just justifies the term coherent space!
}.  In fact, such operators belong to the center $\cal C$ of the von Neumann observable algebra $\overline {\cal U}$ \footnote{Let the algebra ${\cal B}({\cal H})$ of all the linear restricted operators be given over a Hilbert space $\cal H$. Then the topology on ${\cal B}({\cal H})$ that is specified with the seminorms 
$$ p_{\Psi_1,\dots, \Psi_n}^{\Phi_1,\dots, \Phi_n}(A)={\rm max}_{j=1,\dots,n} \vert <\Phi_j, A\Psi_j> \vert$$ is called \cite{Logunov} {\it the weak topology} ($W$-{\it topology}). Here $n$ is a natural number while $\Phi_1,\dots, \Phi_n$ and $\Psi_1,\dots, \Psi_n$ arbitrary vectors in $\cal H$. 
 
 In this terminology, any involutive subalgebra (with the unit element) in ${\cal B}({\cal H})$ closed in the $W$-topology is called {\it the von Neumann algebra} \cite{Logunov}. }.  The totality of all the superselection operators (or merely an arbitrary set of operators generating  $\cal C$) is called the 
{\it superselection rules}.

\medskip The only thing that's obvious is the fact that the magnetic or topological charge is multiple of the unit operator. For instance, in the recent paper \cite{rem1} (see also \cite{Al.S.}), it was shown that  for a typical YM-Higgs (vacuum) configuration  $(A,\Phi)$, a magnetic charge $\bf m$ is a linear
function of the topological charge
\be  
\label{mt1} 
{\bf m} (\Phi,A)= C~ \zeta (\Phi,A), \quad \zeta (\Phi,A)\in {\bf Z}. \ee 

\medskip  But the topological charge  $Q_m$, as that given via (\ref{qum}), is not a local and is not restricted (it takes integers!). Moving on to the terminology of eigenvectors carrying the topologies $n\in \bf Z$, we can write down $$ Q_m |n\rangle = n |n\rangle,\quad n\in{\bf Z}. $$ But $n$'s are not restricted by their modules, and this just does the operator $\hat Q_m$ (with its eigenfunction $Q_m$) not restricted.

\medskip Its will be interesting to speculate, at which circumstances the operator $\hat Q_m$ can be {\it self-adjoint}.  From the point of view of theoretical physics, $\hat Q_m$ is the generator of ("large") gauge transformations with the (nontrivial) topologies at the spatial infinity {\it or} the  conserved charge defined as the integral of the local density  (via $F$ and $A$).  And just in QFT, the conserved charges are implemented as self-adjoint operators  (then $e^{i\alpha Q_m}$ are unitary operators). Now we can choose a domain which is a dense subspace of \linebreak  "good" states possessing finite energies and smooth herewith. The operator $\hat Q_m$ is symmetrical and, at standard conditions,  essentially self-adjoint (i.e. its closure results the self-adjoint operator). 

Now, as we found out at which conditions the operator $\hat Q_m$ can be self-adjoint, we assume that it is so. Then since its spectrum contains the infinite number of integer value without the upper limit, it cannot be restricted: its norm, which is equal to the suprem of the spectrum (taken modulo!), is ${\rm sup}~n \to \infty$.  

\medskip To circumvent the problem that the operator $\hat Q_m$ is not restricted, it is appropriate to introduce the orthogonal projectors
\be \label{project}  P_q = E_{\hat Q_m}({q}), \quad q \in {\bf Z};\quad  P_q P_{q'}=0 ~ {\rm at}~ q\neq q'.\ee

Then, each $P_q$ is a restricted projector. It commutes with each local observable, since the same is correct  for the operator $\hat Q_m$. Really,  the operator $\hat Q_m$, defined via (\ref{magn. zarjad}),  (\ref{qum}), is calculated over the infinitely distant sphere $S_\infty^2$. However, local operators act in a {\it restricted}  domain. 

Further, takes place the natural decomposition

$$ \mathcal H = \bigoplus_q \mathcal H_q, \quad \mathcal H_q = P_q \mathcal H,   \quad q \in {\bf Z}. $$ 
Due to the relation (\ref {project}), the basic property of orthogonal projectors, the Hilbert spaces ${\cal H}_q$ are also mutually orthogonal.  It is obvious that they can be identified with coherent spaces ${\cal H}_\nu$ above.

Herewith in each sector ${\cal H}_q$ , the operator $\hat Q_m$ acts as $qI$  (with $I$ being the unit matrix).

In this sense, the operator $\hat Q_m$ (i.e. {\it the magnetic charge} $m$) implements the superselection in the spire \cite{Logunov}, but not as one restricted operator, multiple the unit, but like a family of projectors extracting the sectors with fixed $Q_m$.

\medskip  
Above, we have give the definition of the von Neumann algebra  ${\cal B}({\cal H})$. Now, after we have shown that the operator $\hat Q_m$ can be realized via a family of projectors $ P_q$  ($q\in {\bf Z}$) extracting the sectors with fixed $Q_m$, i.e. the coherent Hilbert spaces (subspaces) ${\cal H}_q$ inside the Hilbert space $\cal H$, we can assert that the operator $\hat Q_m$, defined  in this manner, belongs indeed to the center  of the von Neumann algebra  ${\cal B}({\cal H})$; in other words, {\it it is a superselection  operator}!

\medskip Note, as a clarification  (see p. 36 in \cite{Logunov}), the projectors $P_q$ set {\it the spectral measure} on ${\bf R}^n$ with its value in the set  $P_q$ of projectors in the Hilbert space $\cal H$.

\bigskip
Returning again to studying the Minkowskian Higgs model with vacuum BPS monopoles quantized by Dirac, note that setting $T\to\infty$ in all the formulas about the topological dynamical variable $N(t)$ seems to be quite correct in the light of the said above about the role of such setting in QFT for free fields as the way to avoid various troubles associated with infrared divergences in quested QFT (in the first place, it is connected closely \cite{Efimov} with the possibility to expand S-matrices by asymptotically free quantum fields at $t\to\pm\infty$ with integrable coefficients).  
\medskip 

The consequence of setting $T\to\infty$ in the Minkowskian Higgs model with vacuum BPS monopoles quantized by Dirac \cite{Dir} is that $\dot N(t)\to 0$ due to (\ref{Poi}).
In other words, angular velocities \cite{Landau1} $\dot N(t)$ of collective solid rotations inside the Minkowskian BPS monopole vacuum suffered the Dirac fundamental quantization \cite{Dir} approach zero in the limit $T\to\infty$. 

And this is, indeed, a bad news. The rotary effects (\ref{rot})-  (\ref{psin}) disappear in this limit $T\to \infty$. In other words, the Minkowskian BPS monopole vacuum will be {\it motionless} as $T\to \infty$.

The only way to resolve a contradiction now arising, is {\it to circumvent the Haag theorem}. It turns out that there are odds to do this, but with numerous warnings. 

For instance, the Haag theorem does not exclude the existence of the interaction (Dirac) picture at violating the translational invariance of QFT by introducing the spatial cut-off \cite{Logunov}. When the interaction  picture exists for the Hamiltonian cutting off in such a way, we are dealing with the so-called {\it local Fock representation} of CCR. However, a difficult mathematical problem arises in this case with the  cut-off removal. 

Such a "local " interpretation of the  interaction  picture, bypassing the Haag theorem, was proposed, for instance, in the work \cite{Guenin}. 

\medskip
In Ref. \cite{fund} there was assumed the so-called "discrete vacuum geometry"
\be \label{RYM}  R_{\rm YM} \equiv SU(2)/ U(1)\simeq {\bf Z}\otimes G_0/U_0,  \ee
with
$$ \pi_1(U_0)= \pi_1(G_0)=0  $$
and 
$$ SU(2)\equiv G; \quad U(1)\equiv U,$$
necessary to justify various rotary effects (in particular, the above described collective solid rotations) proper to the Minkowskian BPS monopole vacuum suffered the Dirac fundamental quantization.

This "discrete vacuum geometry" will be investigated in next papers, which are planed. 

Now we only should like point to the following circumstance just associated with
angular velocities $\dot N(t)$ of collective solid rotations inside the Minkowskian BPS monopole vacuum.

\medskip 



\medskip
Concluding this subsection, we should like discuss some important features of the   Minkowskian Higgs model with vacuum BPS monopoles quantized by Dirac \cite{Dir} that just associated with the properties of the  topological dynamical variable $N(t)$.
\medskip 

First of all, as it follows from (\ref{Poi}),  angular velocities $\dot N(t)$ of collective solid rotations inside the Minkowskian BPS monopole vacuum suffered the Dirac fundamental quantization \cite{Dir} are motion integrals \cite{Pervush1}:
\be \label{moint}
\ddot N(t)=0.
\ee
\medskip

Secondly, as it was shown in Ref. \cite{David3}, introducing  the  noninteger degree of the map $ N(t)$ (in the (\ref{winding num.}) wise) involves the gauge transformations
$${\cal N}_1[{\Phi}^{(0)},N(t)]={\sin[2\pi N(t)]}/{2\pi},
$$
\be\label{cheng1}
{\cal N}_2[N(t)]= \{N(t)-{\sin[2\pi N(t)]}/{2\pi}\};
\ee 
connected with gauge transformations
\be \label{gxN}
X[A^{(n)}]=X[A^{(0)}]+{\cal N}_1[{\Phi}^{(0)},N(t)]+{\cal N}_2[N(t)]=X[A^{(0)}]+N(t). 
\ee 
for the winding number functional $ X[A]$ \cite{rem2,Pervush2}. Herewith
$$ \nu[A]= X[A_{\rm out}]-X[A_{\rm in}]= n(t_{\rm out})-n(t_{\rm in}).  $$
\medskip

Furthermore \cite{Pervush2}, the "boundary" condition (\ref{tbc}) \cite{Pervush3} for the topological dynamical variable $N(t)$ is equivalent to shifts this variable onto integers:
\be
 \label{shift}
 N \Longrightarrow N+n\equiv \tilde N; \quad n= \pm 1,\pm 2,...
\ee 
(as $T\to \infty$ and $t=\pm T/2$).

\medskip
For our further discussion, we should recall the explicit look of Gribov stationary topological multipliers $ v^{(n)}({\bf x})$ in the Minkowskian Higgs model with vacuum BPS monopoles quantized by Dirac.

This proves to be \cite{rem2, LP2,LP1,Pervush2}
\be \label{multipl}
v^{(n)}({\bf x})=
\exp [n\hat \Phi _0({\bf x})],
\ee
with $\hat\Phi _0({\bf x})$ being the {\it Gribov phase}, that is, indeed,
\be
 \label{phasis}
{\hat \Phi}_0(r)= -i\pi \frac {\tau ^a x_a}{r}f_{01}^{BPS}(r), \quad 
f_{01}^{BPS}(r)=[\frac{1}{\tanh (r/\epsilon)}-\frac{\epsilon}{r} ].
 \ee
(where $\tau ^a$ ($a=1,2,3$) are Pauli matrices).

Thus  $\hat\Phi _0({\bf x})$ is an $U(1)\subset SU(2)$ scalar constructed over (topologically trivial) Higgs vacuum BPS monopole solutions \cite{LP2,LP1}.

\medskip
It is easy to see  that Gribov stationary topological multipliers $ v^{(n)}({\bf x})$, (\ref {multipl}), may be replaced with \cite{Pervush2}
\be
 \label{multipl1}
\exp [\tilde N(t)\hat \Phi_0 ({\bf x})],
\ee 
that implies replacing
\be \label{d.v.} \hat A_i^{(N)}= \exp [\tilde N(t)\hat \Phi_0 ({\bf x})] [\hat A^{(0)}_i+\partial _i]\exp [-\tilde N(t)\hat \Phi_0 ({\bf x})] \ee 
for topological Dirac variables $\hat A_i^{D}$ in the initial time instant $t_0$.

This  replacing for topological Dirac variables is quite legitimate since exponential multipliers $\exp(N(t))$ can be always formally included into $U(1)\subset SU(2)$ gauge matrices $u(t,{\bf x})$. 

In this case  topological Dirac variables $\hat A_i^{D}$, (\ref{degeneration}), remain gauge invariant at the replacement (\ref{d.v.}) due to the transformation law \cite{David2, Azimov}
\be
\label{z-n dlja U}
U(t,{\bf x})\to U_u (t,{\bf x}) =  u^{-1}(t,{\bf x}) U(t,{\bf x}),
\ee 
with
$$ U(t,{\bf x})\equiv T \exp \left\{\int  \limits_{t_0}^t d {\bar t}\hat A _0(\bar t,  {\bf x})\right\}.
$$
Thus $ U(t,{\bf x})=0$ as $t=t_0$.

\medskip

Gribov topological multipliers (\ref{multipl1}) may be rewritten in the alternative form \cite{Pervush2}
\be
 \label{spat.as}
 U_{\cal Z}= T \exp [ \int \limits _{t_0} ^t dt'{\hat {\cal Z}}(t',{\bf x})]\vert _{\rm asymptotic}=
\exp [N(t)\hat\Phi _{0}({\bf x})]  \ee  
with 
\be \label{Summ}
{\hat {\cal Z}}(t,{\bf x})\vert _{\rm asymptotic}= {\dot N}(t)\hat\Phi _{0}({\bf x})+O (\frac {1}{r^{l+1}});
\quad l>1. \ee  
Since the topological variable $N(t)$ satisfies the "boundary condition" (\ref{tbc}) \cite{Pervush3} in the initial time instant $t=t_{\rm in}\equiv t_0=-T/2\to-\infty$, in this time instant, Gribov topological  multipliers $ U_{\cal Z}$: (\ref{multipl1}),  (\ref {spat.as}), turn into $ v^{(n)}({\bf x})$ ones at $t=t_0$.

Thus one can assert that in the initial time instant $t=t_0$ there is a natural isomorphism between the sets $\{ U_{\cal Z}\}$ and $\{ v^{(n)}({\bf x})\}$ of Gribov topological  multipliers (the same is correctly, of course, also at $t= T/2$)).

And moreover, Gribov topological  multipliers $ U_{\cal Z}(t_0,{\bf x})$ acts as authomorphisms on the set $\{ v^{(n)}({\bf x})\}$.

The latter fact is in a good agreement with Eq. (\ref{shift}).
More exactly,  the shift (\ref {shift}) of integers $n\in \bf Z$ onto $N(t)$  involves in this case (because of the "boundary condition" (\ref{tbc})) mapping a one (say, "large" \cite{Pervush2}, with the fixed $n\neq 0$) $U(1)\subset SU(2)$ orbit into another, while the alone $U(1)\subset SU(2)$ gauge group remains immovable at this shift.

\medskip
Now there are two important conclusions may be drawn concerning Gribov topological  multipliers $\{ U_{\cal Z}\}$: (\ref{multipl1}),  (\ref {spat.as}).

Firstly, the topological degeneration of Dirac variables $A_i^D$, (\ref{degeneration}), may be reduced, in the initial time instant $t=t_0$, to shifts (\ref {shift}) of the topological variable $N(t)$ onto integers.

Secondly, the us said about Gribov topological  multipliers $ U_{\cal Z}(N(t),{\bf x})$ shows that these multipliers play the role equivalent the role of topological  multipliers $ v^{(n)}({\bf x})$ in the Minkowskian Higgs model with vacuum BPS monopoles quantized by Dirac \cite{Dir}.

In particular, the constraint-shell Hamiltonian of the mentioned model may be expressed in terms of $ U_{\cal Z}(N(t),{\bf x})$ as well as in terms of $ v^{(n)}({\bf x})$, as it was demonstrated in  Ref. \cite{Pervush3}.

In  the papers planed this will be discussed in detail and also the properties of Gribov topological  multipliers $ U_{\cal Z}(N(t),{\bf x})$ will be analysed.
\section{Josephson effect in Minkowskian Higgs model with vacuum BPS monopoles quantized by Dirac. }
 Collective solid rotations inside the Minkowskian Higgs BPS monopole vacuum suffered the Dirac fundamental quantization \cite{Dir} imply the purely real energy-momentum spectrum $P_N$, (\ref{pin}), accompanied by the wave function $\Psi_N$, (\ref{psin}).

Herewith the {\it topological momentum} \cite{LP1} $P_N$, (\ref{pin}), is read easily from the action functional (\ref{rot}).

As it follows from (\ref{pin}), the topological momentum $P_N$ depends explicitly on the $\theta$-angle.

In the present section we attempt to demonstrate our readers that such dependence of $P_N$ on the $\theta$-angle is the particular display of the {\it Josephson effect}, coming to persistent circular motions of material points (quantum fields).

In Ref. \cite{Pervush3} the Josephson effect was explained with the example of two superconductors joined in an electric circuit \footnote{ See the original paper \cite{Josephson} by Josephson. The investigation of B. D. Josephson about tunnelling effects in superconductors were honoured, in 1973y., by Nobel Prize in the sphere of physics.}. 

\medskip
At a contact of two superconductors, the motion of electrons in the both  is strongly correlated. \par
This means that all the electrons possess one and the same momentum and the common wave function \cite{Pervush3} 
\be  \label {contact} \Psi= \prod \limits_{j=1}^n \Psi_j= \exp [ip\sum \limits_{1}^N X_j/\hbar]= \exp[i (pN) \sum \limits_{1}^N X_j/(N\hbar)]\equiv \exp[i {\cal P}{\cal X}/\hbar], \ee
with $\cal X $ being the coordinate of the centre of inertia of  the system of electrons and $\cal P $ being their total momentum. \par
At a contact of two different superconductors: say, in a point $x_0$, a flip of the wave function's phase occurs:
\be  \label {jump} \Psi (x_0+\epsilon) =\exp [i\theta] \Psi (x_0-\epsilon), \quad \epsilon \to 0.
\ee
The origin of latter Eq. is following \cite{Pervush3}.

When a closed (one-dimensional) way is given and $L$ is the length of this closed curve, then points $X$ and $X+L$ coincide and are physically equivalent. \par 
This means that the wave function at the point $X$ coincides, up to the phase $\exp [i\theta] $ ($\vert \theta \vert \leq \pi$), with the wave function at the point $X+L$:
\be  \label {po krygy} \Psi (X+L)= \exp [i\theta] \Psi (X).  \ee 
On the other hand, the fact \cite{Landau3} that any circular motion possesses a discrete momentum spectrum implies that any closed way can be topologically nontrivial (and it is easy to ascertain, applying definite methods of differential geometry, alike those stated in the monograph \cite{Postn4}: in Lectures 3, 4, 26). \par
But also in the analytical way, one can make sure in the nontrivial topological content of circular motions. \par
To do this, it is necessary to utilize the periodicity property of $\exp [i\theta]$ and to express explicitly the momentum $p$ of the considered circular motion in terms of the Planck constant $\hbar$ according to the Heisenberg uncertainty principle \cite{Pervush3}:
\be  \label{impuls}  p=   \hbar ~\frac {2\pi k+ \theta} L, \quad k=\pm (0,1,2,\dots). \ee
Generally, a quantum object (a "{\it quantum train}" in the terminology \cite{Pervush3}) moving along a closed way cannot stop at $\theta \neq 0$, as it follows from (\ref{impuls}). \par 
The  state with the minimum energy at $k=0$ (treated as a "vacuum") corresponds to the {\it persistent motion} of the considered "train" involving the momentum 
\be  \label{persist}
p=\hbar \theta /L.
\ee 
This momentum $p$ of the persistent motion along the closed way disappears in the classical limit $\hbar \to 0$.\par 
This was, in effect, the general mathematical description \cite{Pervush3} of the Josephson effect taking place in physical models involving circular motions.

\medskip
Note that (gauge) physical theories may distinguish by trajectories of  circular motions, and this may tell upon  their specific.\par 
So, for instance, in QED$_{(1+1)}$  \cite{Ilieva1} one deals with circular motions along the circle $S^1\simeq U(1)$ of the infinite radius due to identifying points of the configuration space $\{A_1(x,t)\}$ at the spatial infinity.\par
More exactly \cite{Ilieva1},  the "points"  
\be
\label{stationar1}
 A^{(n)}(x,t)= \exp (i \Lambda ^{(n)}(x)) (A_1 (x,t) + i \frac{\partial_1}{e}) \exp (-i \Lambda ^{(n)}(x)), \quad n \in {\bf Z}, 
\ee 
in the QED$_{(1+1)}$  configuration space $\{A_1(x,t)\}$ \footnote{Here $\Lambda ^{(n)}(x)$ are the $U(1)$ generators, entering explicitly Weyl base elements $P_1^{(n)}\in U(1)$:
$$ P_1^{(n)}(x)= \exp (i \frac{ \Lambda ^{(n)} (x)}{ \hbar}).$$
As it was explained in \cite{Ilieva1},  Weyl base elements $P_1^{(n)}$ would satisfy the spatial asymptotic
$$ \lim _{\vert x\vert \to \pm \infty} P_1^{(n)}(x) =1,       $$
that is equivalent to
$$   \Lambda^{(n)} (\infty)-  \Lambda^{(n)} (-\infty) =2\pi n\hbar,     $$
in order to deal with transverse electric fields $\partial_x \hat E(x,t)=0$ in the model \cite{Ilieva1}.

To satisfy the above spatial asymptotic, it is expediently to assume \cite{Pervush3}
$$  \Lambda ^{(n_\pm)}(x)= \hbar ~2\pi n_\pm \frac {x} {R}       $$
(with $R$ standing for the spatial infinity) at arbitrary values of the spatial coordinate $x$. 

Here, following \cite{Pervush3}, it is convenient to pick out the subsets of positive, $n_+$, and negative, $n_-$, numbers among integers $ n \in {\bf Z}$.} are physically identical.

Herewith under the physical identity (physical equivalence) of the points $ A^{(n)}(x,t)$, with $n$ running about the set $\bf Z$ of integers, in the configuration space $\{A_1(x,t)\}$, the identical probabilities distributions for gauge fields belonging to different  topological domains in the model \cite{Ilieva1} as well as to the one fixed topological domain there may be understood. 
\par
This affects immediately the QED$_{(1+1)}$  wave function $\Psi$: 
\be
\label{faz} 
\Psi (A^{(n+1)})= e^{i\theta} \Psi (A^{(n)}), \quad \vert \theta \vert \le \pi. \ee
Eq. (\ref{faz}) is the particular case of general Eq. (\ref {po krygy}) \cite{Pervush3} in the actual limit $L\to\infty$ for the  infinite large circle $S^1$.

\medskip
In superconductors the Josephson effect becomes possible only at their contact and when a closed electric circuit including these two superconductors is built  \footnote{We recommend our readers the monograph \cite{Blakemore}, \S3.6, for a detailed acquaintance with this question.}.

\medskip
In a liquid helium specimen at rest such persistent motion arises at the spontaneous breakdown of the initial $U(1)$ gauge symmetry of the Bogolubov helium Hamiltonian \cite{N.N.} together with the superfluidity phenomenon  \cite{Landau}. 

The specific of a liquid helium specimen at rest model \cite{N.N., Landau} is such that superfluid potential motions (proceeding with velocities do not exceeding a crucial ${\bf v}_0$ one \cite{rem1, Landau}) coexist there with vortices. 

A good analysis of vortices in a liquid helium specimen (at rest) was performed in the monograph  \cite{Halatnikov}, in \S\S ~ 30, 31.

It turns out herewith that that nontrivial topologies $n\neq 0$ in a liquid helium specimen (at rest) just correspond to vortices therein, while the trivial topology $n=0$ corresponds to the superfluidity phenomenon 
\cite{rem1,N.N.,Landau}.

More exactly, there is the simple relation \cite{Halatnikov} associated the tangential velocity ${\bf v}^{(n)}$ of a rectilinear vortex in a liquid helium (at rest) to the given topological number $n\in \bf Z$:
\be \label{vixr}
n= \frac{m}{2\pi \hbar} \oint \limits _\Gamma {\bf v}^{(n)}d {\bf l}, \ee  
with $m$ being the mass of the helium atom; $ d {\bf l}$ being the element of the length along the axis $z$ of this (rectilinear) vortex.

It follows from Eq.  (\ref{vixr}) that this cyclic integral vanishes as $n=0$.
\par
On the other  hand \cite{rem1,Landau}, it is mathematically equivalent to 
$$ {\rm rot}~ {\bf v}^{(0)}=0,$$
with the velocity vector ${\bf v}^{(0)}$ chosen to be directed along the vector ${\bf v}_0$ of the superfluid motion in the liquid helium. 
\par
Thus the trivial topology $n=0$ corresponds really to the superfluid potential motion, without friction forces between the liquid helium specimen and walls of the (rested) vessel where it is contained. \par
Moreover, to the same result one comes by setting the (semi)classical limit $\hbar\to 0$ in (\ref{vixr}):  
\be \label{semiclassic}
m \oint \limits _\Gamma {\bf v}^{(n)}d {\bf l}=2\pi n ~0=0.
\ee 
On the other hand, according to the Landau two-component phenomenological model of liquid helium II \cite{Landau} and its "quantum version" \cite{N.N.}, created by N. N. Bogolubov with 
co-authors, superfluidity in a helium is possible only in the limit $p\to 0$ of small transferred momenta, i.e. when the momentum $m {\bf v}_0$ of the superfluid motion in a liquid helium II specimen is comparable by its absolute value with momenta $p_{\rm ph}\to 0$ of phonons, belonging  to the excitations spectrum in this specimen. \par
In turn \cite{Halatnikov}, the part of the energy spectrum in a liquid helium corresponding to phonons is set by Eq.
$$ \epsilon \sim c_{s}~p_{\rm ph},$$
with $ c_{s}$ being the sound velocity. \par
In this case of small transferred momenta, the semi-classical limit $\hbar\to 0$ would correspond to large sizes (may be assumed to be infinite) of the vessel where the liquid helium specimen is contained in order for  this specimen to posses the manifest superfluidity: it is true due to the Heisenberg uncertainty principle.  \par
\medskip
Thus one can assert that the Josephson effect \cite{Pervush3} in a liquid helium II specimen comes to (rectilinear) vortices (\ref{vixr}) interspersing the topologically trivial superfluid  component in the liquid helium and bearing  traces of nontrivial topologies $n\in{\bf Z}$ \footnote{It may be argued that the coexistence of superfluid motions and (rectilinear) vortices in a liquid helium II specimen testifies in favour of the first-order phase transition 
occurring  there}. \par
Geometrical spaces of these (rectilinear) vortices in a liquid helium \cite{Halatnikov} are, indeed, infinite narrow cylinders. \par 
The important specific of the Josephson effect in a liquid helium II specimen is that the 
$\theta$-angle can and would attain its zero in the liquid helium II at rest theory \cite{Halatnikov} simultaniously with $n=0$.\par
It is associated again with the superfluidity claims \cite{N.N.,Landau} would be imposed undoubtedly onto the liquid helium II at rest theory \cite{Halatnikov}.\par 
In this case rotary effects disappear in the zero topological sector of the model \cite{Halatnikov} together with circular integrals of rotary velocities ${\bf v}^{(n)}$ ($n\in {\bf Z}$); this just corresponds to setting $\theta=0$ for this sector \cite{Pervush3}. \par 

\bigskip
Considering  the examples of QED$_{(1+1)}$  \cite{Ilieva1} and the liquid helium II at rest theory \cite{rem1,N.N.,Landau, Halatnikov} with their specifics of the Josephson effect, now we can proceed to studying the features of this effect in the Minkowskian Higgs model with vacuum BPS monopoles quantized by Dirac \cite{Dir}.

\medskip
First of all, it is obvious that Eq. (\ref{pin}) for the topological momentum $P_N$ of the Minkowskian BPS monopole vacuum (suffered the Dirac fundamental quantization) is also the particular case of general Eq. (\ref {impuls}) \cite{Pervush3}.

Herewith one would set (conditionally) $L=1$ in (\ref {impuls}) in order to get (\ref{pin}).

\medskip
Further, Eq. (\ref {po krygy}) \cite{Pervush3} for the wave function $\Psi$ is transformed in the mentioned Minkowskian Higgs model into Eq. \cite{David2}
\be 
\label{per}
\Psi_{N} (N+1)=e^{i\theta}\Psi_{N}(N).
\ee 
This Eq. is analogous to Eq. (\ref {faz}) in (Minkowskian) QED$_{(1+1)}$  \cite{Ilieva1}. 

\medskip
Finally, as it was already discussed in Introduction, the Josephson effect in the \linebreak Minkowskian Higgs model with vacuum BPS monopoles quantized by Dirac may be reduced \cite{David2}
to the presence of vacuum "electric" monopoles $E_i^a$, (\ref{sem}).

Herewith in the zero topological sector of this Minkowskian Higgs model any vacuum "electric" monopole $E_i^a$ achieves its nonzero (until $\theta\neq 0$) minimum (\ref{se}) \cite{David2} \footnote{ We oughtn't  be afraid of the $\epsilon^{-1}$ dependence (with $\epsilon\to 0$ as $V\to \infty$) in Eq. (\ref{se}) since just this yields the $O(\epsilon)$ convergence of the free rotator action (\ref{rot}) \cite{David2}}.

Indeed, Eq. (\ref{se}) is the particular case of  Eq. (\ref {persist}) \cite{Pervush3} (at setting $L=1$).

It is associated with the natural interpretation \cite{Gitman} of  various "electric" fields (if nonzero) in all kinds of non-Abelian theories as canonical momenta conjugate to temporal components $A_0$ of gauge fields.

In particular, a vacuum "electric" field $E_i^a$, (\ref{se}), plays the role of the canonical momentum conjugate to a vacuum potential $Z^a$, (\ref {zero}).

\medskip
Eq. (\ref{sem}) allows to assert \cite{David2} that    "persistent field motions" around the "cylinder" of the diameter $\sim \epsilon$, with its symmetry axis coinciding with the axis $z$ of the chosen (rest) reference frame,  take place,  recognized as the Josephson effect \cite{Pervush3} occurring in  the Minkowskian physical non-Abelian BPS monopole vacuum suffered the Dirac fundamental quantization \cite{Dir}.

These   "persistent field motions" around the infinitely narrow "cylinder" of the diameter $\sim \epsilon$  come to  solid rotations inside the Minkowskian physical non-Abelian BPS monopole vacuum. 

And moreover, Eq. (\ref {per}) (as the particular case of general Eq. (\ref {po krygy}) \cite{Pervush3}) for the wave function $\Psi_N$ of the Minkowskian physical non-Abelian BPS monopole vacuum suffered the Dirac fundamental quantization describes correctly the Josephson effect in that vacuum with the above discussed "cylinder" topology (in the terminology \cite{Pervush3}) 

\medskip 
Thus \cite{David2,Pervush3} a field theoretical analogy of the Josephson effect is available  in the Minkowskian Higgs model with vacuum BPS monopoles quantized by Dirac \cite{Dir}: a 
circular current without sources \footnote{In studying planed, the interconnection between "persistent field motions" around the infinitely narrow "cylinder" of the diameter $\sim \epsilon$ and the "discrete" vacuum geometry just assumed for justifying solid rotations inside the Minkowskian physical non-Abelian BPS monopole vacuum suffered the Dirac fundamental quantization \cite{Dir} will be revealed.}.

\medskip 
Note that in (Minkowskian) QED$_{(1+1)}$  \cite{Ilieva1}, there is an analogue of  vacuum "electric" monopoles (\ref{se}) inherent in the Minkowskian Higgs model with vacuum BPS monopoles quantized by Dirac.

It is the electric field  \cite {Ilieva1,Coleman, Gogilidze}
\be
\label{Col.sp}
E\equiv G_{10}=\dot N[A](t) \frac {2\pi}{e} =e(\frac {\theta}{2\pi}+k).
\ee 
In the zero topological sector of QED$_{(1+1)}$  \cite{Ilieva1}, i.e. when $k=0$ \cite{Pervush3}, 
\be \label{never vanish}
E_{\rm min}= \frac{e\theta}{2\pi}.   
\ee
Herewith the topological dynamical variable $N[A](t)$ inherent in QED$_{(1+1)}$  \cite{Ilieva1} possesses the properties similar to these $N(t)$ possesses in the Minkowskian Higgs model with vacuum BPS monopoles quantized by Dirac \cite{Dir}.

For instance, $N[A](t)$ may be represented as \cite{Pervush3}
\be\label{Pontrjaginp}
N[A](t)= (n_--n_-) \frac t T +(n_+ +n_-)\frac 1 2, \ee
by analogy with (\ref{tbo}) (with that distinction that in QED$_{(1+1)}$ it was assumed  \cite{ Pervush3} $n_-<0$ and $n_+>0$, as we have mentioned above).

Then again the "boundary condition"
\be\label{bond.cond}
N[A](\pm T/2)\to n_\pm \vert_{\vert x \vert \to R/2}. 
\ee
(where $T\to \infty$ and $R\to \infty $ is set \cite{Pervush3, Ilieva1}) is satisfied by the topological dynamical variable $N[A](t)$. 

Just the topological dynamical variable $N[A](t)$ determine in QED$_{(1+1)}$  \cite{Ilieva1} circular motions around the circle $S^1\simeq U(1)$ of the infinite radius,  that is the essence of the Josephson effect \cite{Pervush3}
in this model.

These circular motions may be described with the aid of the action functional \cite{Pervush3}
\be
 \label{twoac}
 S (R,T,  \nu ) =   \int \limits _{-T/2} ^{T/2} L(t)dt,  \quad  L(t)= \frac{1}{2V} ( \frac{2\pi}{e} )^2 \dot N[A] ^2 (t)  \equiv \frac{1}{2} M  \dot N[A]  ^2, 
\ee 
 with $V$ being the spatial volume,
$$ V \equiv  \int\limits _{-R/2} ^{R/2} dx =R,$$
and
\be \label{dotn}  \dot N[A] (t)= \nu/T    \ee 
due to (\ref{bond.cond}) (where again $\nu=n_+-n_-$).
\section{Conclusion}
In the present study, we have discussed the nontrivial topological dynamics inherent in the Minkowskian Higgs model with vacuum BPS monopoles quantized by Dirac  \cite{Dir}.

This dynamics, coming to the specific Josephson effect \cite{Pervush3}  in the enumerated model, i.e. to the existence of  collective solid rotations inside the {\it physical} BPS monopole vacuum.

These solid rotations are described correctly by the free rotator action 
functional (\ref{rot}) \cite{David2}, implicating the topological dynamical variable $N(t)$ (given via Eq. (\ref {winding num.}) \cite{Pervush2}), vacuum "electric" monopoles $F_{i0}^a$ (given via Eq. (\ref {se}) \cite{David2}) and the purely real energy momentum spectrum  (\ref{pin}).

They also don't vanish at nonzero values $\theta\neq 0$ of the $\theta$-angle, while "geometrically", there are, indeed, rotations around the infinitely narrow cylinder of the effective diameter $\sim \epsilon$ (with $\epsilon$ being the typical size of BPS monopoles) along the axis $z$  of the chosen (rest) reference frame. 

These results about the nontrivial topological dynamics inherent in the Minkowskian Higgs model with vacuum BPS monopoles quantized by Dirac are got at the Gauss-shell reduction of that model in terms of topological Dirac variables $A^D_i$ ($i=1,2$), (\ref{degeneration}), gauge invariant and transverse functionals of YM fields.

As a result of this constraint-shell reduction, the family $Z^a$, (\ref {zero}), exist satisfying the Gauss law constraint (\ref {homo}): the second-order homogeneous differential equation in partial derivatives.

This picture of the collective solid rotations inside the BPS monopole vacuum (suffered the Dirac fundamental quantization \cite{Dir}) seems to be correct at least at the absolute zero temperature $T=0$,  when these  rotations proceed in the "non-stop" regime \cite{Pervush3} and "friction forces" between this BPS monopole vacuum and its surroundings are absent.

\medskip
In Ref.  \cite{fund}, it was pointed out that grounding the above described nontrivial topological dynamics in the Minkowskian Higgs model with vacuum BPS monopoles quantized by Dirac lies in assuming the "discrete" vacuum geometry (\ref{RYM}) for the appropriate vacuum manifold (degeneration space) $R_{\rm YM}$.

This idea will be us developed in the further study. 

In particular, series of interesting properties of the discrete vacuum manifold $R_{\rm YM}$, (\ref{RYM}),  will be revealed and discussed.

It will be argued (continuing the job begun in \cite{fund}) the presence of three kinds of topological defects inside this manifold: thread and point hedgehog topological defects inside $R_{\rm YM}$.

We shall see that just thread topological defects inside the discrete vacuum manifold $R_{\rm YM}$, (\ref{RYM}), are the cause of various rotary phenomena in the Minkowskian Higgs model with vacuum BPS monopoles quantized by Dirac \cite{Dir}.

It is, in fact, the same mechanism that causes (rectilinear) vortices \cite{Halatnikov} in a liquid helium II specimen.

\medskip

The existence of thread topological defects inside the vacuum manifold $R_{\rm YM}$, (\ref{RYM}), is associated with the presence therein of rectilinear threads around which collective solid rotations (as a specific kind of vortices) proceed inside the BPS monopole vacuum suffered the Dirac fundamental quantization \cite{Dir}.

In the study planed, it will be shown, repeating the arguments \cite{Al.S.}, that there are YM fields 
$$ A  _\theta(\rho, \theta, z) = A_\mu \partial x^\mu/ \partial \theta $$ 
These fields may be always represented as \cite{Al.S.}
$$  A  _\theta (\rho, \theta, z)=   \exp(iM\theta) A  _\theta (\rho) \exp(-iM\theta),    $$ 
with $M$ being the generator of the group $G_1$ of global rotations compensating changes in the vacuum (Higgs-YM) configuration $(\Phi^a,A_\mu^a)$ at rotations around the  axis $z$ of the chosen (rest) reference frame. 

The elements of $G_1$ may be set as \cite{Al.S.}
$$ g_\theta =\exp(iM\theta).    $$
YM fields $A_\theta $ are manifestly invariant with respect to shifts along the axis $z$.

Note that rectilinear threads $A_\theta $ don't coincide with vacuum YM BPS monopole solutions, and, on the contrary, there are, indeed, gaps between directions of vectors ${\bf B}_1$:
$$ \vert {\bf B}_1 \vert \sim \partial _\rho A_\theta (\rho,\theta, z), $$ 
and $\bf B$, given by the Bogomol'nyi equation (\ref{Bog}).

These gaps testify in favour  the first-order phase transition occurring in the Minkowskian Higgs model with vacuum BPS monopoles quantized by Dirac \cite{Dir}, where the "magnetic" field $\bf B$, given via the Bogomol'nyi  equation (\ref{Bog}) \cite{rem1, LP2,LP1,Al.S.}, may be treated as the order parameter in the enumerated  model.

As it was demonstrated in the original papers \cite{BPS}, this "magnetic" field $\bf B$ diverges as $r^{-2}$ at the  origin of coordinates, and it is always a sign of a phase transition (either of the first or the second order) occurring in the Minkowskian Higgs model with vacuum BPS monopoles. 

\medskip 
Unlike YM modes, suffered gaps near the axis $z$, Higgs vacuum BPS  monopole modes may be continued in the (smooth) wise in this spatial region. In detail, the situation is following.

As it was demonstrated in Ref. \cite{David3} (see also \cite{rem2}), the Higgs BPS ansatz 
$$ f_{01}^{BPS}(r)=[\frac{1}{\tanh (r/\epsilon)}-\frac{\epsilon}{r} ]        $$
has the asymptotic  
$$ f_{01}^{BPS}(0)=0; \quad f_{01}^{BPS}(\infty)=1.
 $$
On the other hand, as it was shown in the monograph \cite{Al.S.}, there exist z-invariant (vacuum) Higgs solutions in a (small) neighbourhood of the origin of coordinates:
$$  \Phi^{(n)} (\rho, \theta, z)= \exp (M\theta)~ \phi  (\rho) \quad (n\in {\bf Z}),     $$
can join vacuum Higgs BPS monopoles, belonging to the same topology $n$ and disappearing \cite{David3} at the origin of coordinates, in a (smooth) wise.

Herewith, speaking "in a smooth wise", we imply that the covariant derivative $D\Phi$ of any vacuum Higgs field $\Phi_a^{(n)}$ merges with the covariant derivative of such a vacuum Higgs BPS monopole solution. \par

This requirement for vacuum Higgs fields $\Phi_a^{(n)}$ to be smooth is quite natural if the goal is pursued, in the Minkowskian Higgs model with vacuum BPS monopoles quantized by Dirac \cite{Dir}, to justify various rotary effects inherent in this model.

In particular, vacuum "electric" monopoles (\ref{se}) are directly proportional to \linebreak $D_i (\Phi_k^{(0)})~ \Phi_{(0)}$. 

These vacuum "electric" monopoles, in  turn, enter explicitly the action functional (\ref{rot}), describing, in the Dirac fundamental quantization scheme \cite {Dir}, collective solid rotations inside the Minkowskian BPS monopole vacuum.

\medskip
Moreover, as it will be discussed in the next study, such (smooth) sawing together  appropriate vacuum Higgs modes $\Phi^{(n)}$ and BPS monopoles is intended to remove the following problem.

As it was discussed recently in the papers \cite {rem1, rem2}, manifest superfluid properties of the Minkowskian BPS monopole vacuum (suffered the Dirac fundamental quantization \cite {Dir}) are set by the Bogomolny'i, (\ref{Bog}), and Gribov ambiguity,
$$   [D^2 _i(\Phi _k^{(0)})]^{ab}\Phi_{(0)b} =0,    $$
equations.

Mathematically, latter Eq. is the consequence of the Bogomolny'i equation (\ref{Bog}) and the Bianchi identity $D~B=0$.

The both above Eqs. describe correctly \cite {rem1, rem2} the Minkowskian BPS monopole vacuum suffered the Dirac fundamental quantization as a superfluid potential (incompressible) liquid.

Thus one can assert \cite {Pervush1} that
$$ D~B\sim D~E =0  $$
for vacuum "magnetic" and "electric" tensions in the quested Minkowskian Higgs model, i.e. that these tensions are, indeed, "transverse" vectors collinear each other.

Note that Eq. (\ref{se}) \cite{David2} just reflects this collinearity. 

This implies, on the face of it, a   contradiction  between collective solid rotations inside the Minkowskian BPS monopole vacuum suffered the Dirac fundamental quantization \cite {Dir} and its manifest superfluid and potential nature \cite {rem1, rem2}, excluding formally any rotations (due to the same reasoning as  for the superfluid component in a liquid helium II specimen \cite{N.N.,Landau}).

Going out from this contradiction seems to be just in locating (topologically nontrivial) threads in the infinitely narrow cylinder of the effective diameter $\epsilon$ around the axis $z$ and in  joining (in a smooth wise) vacuum Higgs fields $\Phi_a^{(n)}$ and BPS monopole solutions.

In this case collective solid  rotations (vortices) inside the Minkowskian BPS monopole vacuum, occurring actually in that spatial region around the axis $z$ and described correctly by the action functional (\ref{rot}), become quite "legitimate", and simultaneously, the Gauss law constraint (\ref{homo})  is satisfied outward  this  region with smooth vacuum "electric" monopoles $E_i^a$ \cite{David2},  (\ref{se}).

\medskip
The said indicates the coexistence of two thermodynamic phases inside the Minkowskian BPS monopole vacuum suffered the Dirac fundamental quantization \cite {Dir}, i.e. the first-order phase transition occurring therein (additional to the second-order one associated with the spontaneous breakdown of the initial $SU(2)$ gauge symmetry down to the $U(1)$ one).

There are the thermodynamic phases of collective solid  rotations and superfluid potential motions inside that physical vacuum. 

Herewith the enough clear-cut picture can be observed how the enumerated thermodynamic phases are distributed inside the discrete vacuum manifold $R_{\rm YM}$, (\ref{RYM}).

Thread topological defects (vortices), associated with rectilinear threads $A_\theta$, are located intimately near the axis $z$  of the chosen (rest) reference frame. Actually, they refer to the cylinder of the effective diameter $\epsilon$ with $z$ serving its symmetry axis.

Simultaneously, superfluid potential motions refer to the spatial region out of this cylinder, including the spatial region $\vert {\bf x}\vert \to\infty$ (corresponding to the infrared region of the momentum space).

The question about the boundary between these phases (and surface effects associated with this boundary) is, however, not simply one. This question (we leave for later studies) is complicated, for instance, by the discrete geometry of the vacuum manifold $R_{\rm YM}$.

\medskip
The important consequence (will be us studied in one of works what follow) of the presence of rectilinear threads $A_\theta $ in the Minkowskian Higgs model with vacuum BPS monopoles quantized by Dirac and involving the "discrete" vacuum geometry (\ref{RYM}) is the effect 
\cite{Al.S.} of annihilating two equal magnetic charges ${\bf m}_1= {\bf m}_2={\bf m}(n)\neq 0$ ($n\in{\bf Z}$) colliding at crossing a rectilinear topologically nontrivial thread $A_\theta (n)$.

Note  that this effect takes place actually within a fixed topological domain inside the discrete vacuum manifold $R_{\rm YM}$, (\ref{RYM}), possessing the topological number $n$. 

Colliding magnetic charges with different topological numbers can be suppressed in the spatial region near the axis $z$ (of the chosen rest reference frame).

The said means the possible annihilation of all the topologically nontrivial YM vacuum BPS monopole modes and excitations over  this BPS monopole vacuum (suffered the Dirac fundamental quantization \cite{Dir}) during a definite time.

As a consequence of such possible annihilation, Higgs (BPS monopole) modes should be free electric fields: their electric charges $e$ which
 are dual (due to the Dirac quantization \cite{Dirac} of  electric and magnetic charges) to zero  magnetic charges can only  survive upon the above described annihilation.

Such situation when Higgs modes possess arbitrary electric charges $e$, while magnetic charges ${\bf m}\neq 0$ are confined is referred to as  the {\it Higgs phase} in modern physical literature (see e.g. \cite{Hooft}) \footnote{Herewith it is understood customary (for instance, in Ref. \cite{Hooft}) that  all magnetic charges are confined by (narrow) Meisner flux tubes, similar to ones in a  superconductor \cite{Cheng}.

  In  turn, this involves the linearly  increasing "Mandelstam" potential $O(Kr)$ (with $K$ being the string tension) between YM monopole and antimonopole.

In the Minkowskian Higgs model with vacuum BPS monopoles quantized by Dirac, that is the subject of our discussion, the role of the  "Mandelstam" mechanism \cite{Cheng} annihilating magnetic charges via their confinement by Meisner flux tubes diminish (perhaps, only at finite temperatures $T\neq 0$, a contribution from the linearly  increasing "Mandelstam" potential $O(Kr)$ in that model is possible, with the specific shape $O[K (g^2T)^{-1}]$ \cite{Linde}).

This temperature-depending "Mandelstam" potential can and would join a linear combination of the Coulomb $r^{-1}$ and "golden section" potentials \cite{LP2,LP1,Pervush2,David2,fund} proper to the Minkowskian Higgs model with vacuum BPS monopoles quantized by Dirac.

On the other hand, at $T\to 0$, the role of the above described mechanism \cite{Al.S.} annihilating magnetic charges ${\bf m}\neq 0$ via their colliding with topologically nontrivial threads $A_\theta$ increases in forming the Higgs phase inside the BPS monopole vacuum, supplanting actually the "Mandelstam" annihilation mechanism \cite{Cheng}.

It is, obviously the merit of assuming the "discrete" geometry (\ref{RYM}) for the vacuum manifold $R_{\rm YM}$, calling to justify the Dirac fundamental quantization of the Minkowskian Higgs model with vacuum BPS monopoles. }.

\medskip
If quarks are incorporated in the Minkowskian Higgs model with vacuum BPS monopoles quantized by Dirac \cite{Dir}, disappearing topologically nontrivial YM modes via the "colliding" mechanism \cite{Al.S.} can cause the possibility to observe free "coloured" quarks in the spatial region along the axis $z$ and wherein near the origin of coordinates (in light of the said above).

This can serve as a (perhaps, enough rough) representation for the asymptotical freedom of quarks  in that model.




The important gain we got in the present study is also that, at the fixed spatial volume $V$, the 'effective' mass  $m/\sqrt \lambda$ is directly proportional to the coupling constant $g$, i.e., together with $g$, obeys the    Callan $-$ Symanzik equation \cite{CS}.

\begin{appendix} 
\section{Appendix 1. Specific of asymptotic states in Euclidian instanton theory.}   
\renewcommand{\theequation}{A.\arabic{equation}}
\setcounter{equation}{0}
The specific of the Euclidian instanton YM theory \cite{Bel} is associated, in the first place, with the Wick rotation $t=-x_4$ of the time axis.

In   this case the finite action functional \cite{Pervush2,Cheng}
\be \label{Evd}
S_{\rm Eucl} (A)=\frac{8\pi^2}{g^2}\nu,
\ee  
where now the Pontryagin index $\nu$ is expressed through instantons $A$ in the standard way (\ref{Ch-S}), (\ref {wind}) \cite{Pervush2} with fixing the {\it Weyl} gauge $A_0=0$ \cite{Cheng} for temporal components of these instantons.

It is assumed (and Eqs. (\ref{Ch-S}), (\ref {wind}) indicate this fact) the topological degeneration of the Euclidian instanton vacuum \cite{Bel}, consisting herewith of purely gauge stationary configurations \cite{Pervush2,Cheng}
\be \label{cl.vac} {\hat A}_i\Rightarrow L^n_i\equiv v^{(n)}({\bf x})\partial_i v^{(n)}({\bf x})^{-1} \quad {\rm as}~ \vert {\bf x}\vert \to\infty; \quad v^{(n)}({\bf x})\in SU(2).\ee
This conception of the topologically degenerated Euclidian instanton vacuum was discussed in Refs. \cite{Bel}  and \cite{Pol} \footnote{Generally speaking, the topological degeneration is proper to vacua in gauge theories independently on the space: either the Euclidian or the Minkowski one, in which these gauge theories are considered.}.

\medskip
When the Euclidian time $x_4$ runs from $x_4=-\infty$ to $x_4=+\infty$, an instanton $A$ interpolates between the classical vacua (\ref{cl.vac}) with the topological numbers $n_+$ and $n_-$, respectively (as it was suggested for the first time in some papers by V. N. Gribov). Herewith this instanton $A$ possesses the topological number $\nu= n_+-n_-$.

It is well known that the mentioned transitions between classical vacua with different topological numbers (they refer to  $\vert x\vert\to\infty$, in the good agreement with our discussion in Subsection 2.2 about the crucial role in QFT of asymptotical states at $t\to\pm\infty$) proceed with the complete energy $\epsilon=0$.

This corresponds (see e.g. the analysis of the Euclidian instanton YM model \cite{Bel} in Refs. \cite{Cheng, Lenz}) to the asymptotical condition imposed onto the YM tension tensor $ F_{\mu \nu}^a$ (with group indices $a$ omitted)
\be \label{tens.a}
F_{\mu \nu}(x)\to 0, \quad \vert {x} \vert \to \infty, 
\ee
that is mathematically equivalent to the asymptotic
\be \label{EB}
\vert {\bf B}\vert =\vert {\bf E}\vert \to 0, \quad \vert {x} \vert \to \infty
\ee
for the "magnetic" tension $\bf B$ and the "electric" tension $\bf E$.

Moreover, going over to the Euclidian space $E_4$ from the Minkowski one in the instanton theory \cite{Bel}, implicating the time coordinate $x_4\equiv \tau=it$, involves also semi-classical paths between  classical vacua (\ref{cl.vac}) "in" and "out" (respectively, at $x=\mp \infty$).

In this case transition amplitudes between the mentioned "in" and "out" vacua turn out having the look \cite{Cheng, Coleman1}
\be 
\label{am}
T\sim e^ {-S_{\rm Eucl}/\hbar}[1+ O(\hbar)].
\ee
\medskip
On the other hand, going over to the Minkowski space in the instanton non-Abelian model is not desirable.

In that case  definite problems arise.

For instance, instantons turn into complex fields in the Minkowski space, although  they are real  in the Euclidian space $E_4$.

It is necessary to recall herewith that in the Euclidian space $E_4$, instantons \cite{Bel} satisfy the duality conditions \cite{Cheng}
\be \label{dual}
F_{\mu \nu}=\pm \tilde F_{\mu \nu}; \quad \tilde F_{\mu \nu}= \frac 1 2 \epsilon^{\mu \nu \alpha\beta} F_{\alpha\beta};
\ee 
that is equivalent to \cite{Pervush1,Arsen}
\be \label{dual1}
B=\pm E
\ee
(where $B$ is the absolute value of the "magnetic" field and $E$ is the absolute value of the "electric" field)
for the YM tension tensor $ F_{\mu \nu}$.

Indeed, when the duality conditions (\ref{dual}) are satisfied, the appropriate (Euclidian) action functional
\be
\label{aYM} S_{\rm Eucl}=\int d^4 x \frac {1}{2g^2} ~{\rm tr} ~   (F_{\mu \nu} F^{\mu \nu}) \ee 
achieves its minimum, as it was demonstrated in \cite{Cheng}.

The mathematical equivalence between Eqs. (\ref{aYM}) and (\ref {Evd}) for the Euclidian action functional may be checked easy repeating the arguments \cite{Cheng}, however we omit  this checking here.

Herewith the alone Hodge duality operation $*$, specified in the (\ref{dual}) way, takes only two eigenvalues,
\be \label{Hodge}
*^2 =1, \quad *=\pm 1, \ee 
in the Euclidian space $E_4$.

But  in the Minkowski space the Hodge duality operation $*$ is specified badly (as it may be shown, repeating the arguments \cite{Postn4, Postn2}). This just serves the cause why instantons turn into complex fields in the Minkowski space. 

\medskip
The topological degeneration of the Euclidian instanton (purely gauge) vacuum and the existence of semi-classical paths (tunnelling transitions) of the shape (\ref {am}) \cite{Cheng,Coleman1}
 between asymptotical "in" and "out" states with  the total energy $\epsilon=0$ allow to introduce the so-called $\theta$-vacuum as a specific superposition of classical vacua (\ref{cl.vac}).

Following  \cite{Cheng},  such superposition may be written down as
\be \label{decomposition} \vert\theta> =\sum \limits_n e ^{-in\theta}\vert n>.
\ee
Herewith the parameter $\theta$ is referred to as the $\theta$-angle (or as the \it  Bloch pseudomomentum\rm) in the modern physical literature.

It is relevant to consider now the wave function $\Psi_0[A]$ that is indeed the eigenfunction of the Hamiltonian \cite{Pervush1}
\be 
\label{HYM} 
H(E,A)= \int d^3x \frac {1}{2}[(E_i^a)^2+ (B_i^a)^2], \ee  \be \label{E} E_i^a= \partial_0 A_i^a, \quad  B_i^a= \frac {1}{2} \epsilon_{ijk} (\partial^j A^{ak} -\partial^k A^{aj} +\frac {g}{2}\epsilon ^{abc}A_{b}^j A_{c}^k), 
\ee
disappearing over the remote surface $\vert {\bf x}\vert \to\infty$ because of the asymptotic (\ref{EB}) \cite{Lenz}.

Thus on the quantum level, the Schr\"odinger equation \cite{Pervush1}
\be 
\label{Schrod} 
{\hat H}(i\delta/\delta A, A) \Psi _0[A]=0; \quad {\hat H}= \int  d^3x \frac {1}{2} [E^2+B^2]; \quad E= \frac {\delta}{i\delta A}; 
\ee
takes  place.

\medskip
The action of the $\theta$-vacuum (\ref {decomposition}) onto the wave function $\Psi_0[A]$ comes to \cite{Cheng}
\be 
\label{period} 
T_1 \Psi _0[A] = e^{i\theta}\Psi _0.
\ee 
Herewith the operator $T_1$, called the {\it raising operator} \cite{Cheng}, involves the winding number 1 as its eigenvalue: 
\be 
\label{rise} 
T_1 \vert n>=  \vert n+1>,
\ee 
or, in terms of the winding number functional $X[A]$, (\ref{wind}), as \cite{LP1}
\be \label{tx}
T_1~X[A]=X[A]+1. 
\ee

\medskip
In the series of physical literature, e.g.   \cite{Pervush1,Arsen, Galperin} (see also Ref. \cite{fund}), the analysis of the instanton Euclidian $\theta$-vacuum was performed and additional properties of this vacuum were discovered.

In particular, the properties of the instanton Euclidian $\theta$-vacuum at arbitrary values $\epsilon$ of the energy-momentum spectrum were investigated.

This proved to be very helpful, although it is enough to consider only the $\epsilon=0$  value of this spectrum, corresponding to the boundary conditions (\ref {tens.a}), (\ref{EB}) \cite{Cheng,Lenz} imposed onto instanton YM fields.

For an arbitrary  $\epsilon$, the (instanton) Euclidian $\theta$-vacuum may be specified, following Ref. \cite {Callan1}, as the ground  state satisfied  the equations \cite{Pervush1,Arsen, Galperin} 
\be
\label{Schrodinger}
{\hat H}(i\delta/\delta A, A) \Psi _\epsilon[A]=\epsilon \Psi _\epsilon[A],  \ee 
\be 
\label{ve} 
\nabla _i^{ab}(A) (\frac {\delta}{i\delta A_{i b}}\Psi_\epsilon [A])=0; \quad \nabla _i^{ab}(A)= \delta ^{ab}\partial_i- g\epsilon ^{abc}A_{ic}; 
\ee   
and 
\be 
\label{period1} 
T_1 \Psi _\epsilon [A]= e^{i 1\cdot \theta} \Psi_\epsilon[A].
\ee 
Herewith the Schr\"odinger equation (\ref{Schrodinger}) generalizes the one (\ref {Schrod}), while Eq. (\ref {period1}) generalizes Eq. (\ref {period}).

Eq. (\ref{ve}) reflects \cite{Pervush1} the invariance of the wave function $\Psi _\epsilon [A]$ (the eigenfunction of the $\theta$-vacuum) with respect to "small" $SU(2)$ gauge transformations belonging to the trivial topology $n=0$.

Eq. (\ref{ve})  may be "translated" to the claim \cite{Arsen}  that the "electric" field $\bf E$ is transverse (formally):
\be 
\label{Etrans}
\nabla _i E^i\Psi _\epsilon [A]=0.
\ee
Indeed, the vector $\bf E$ disappears over an infinitely remote surface $\vert {\bf x}\vert \to\infty$ due to the boundary conditions (\ref {tens.a}), (\ref{EB}) imposed onto instanton YM fields.

Unlike (\ref{ve}), associated  with "small" $SU(2)$ gauge transformations, Eq. (\ref {period1}) is associated  with "large" $SU(2)$ gauge transformations belonging to  nontrivial topologies $n\neq 0$.

Herewith the eigenfunction $\Psi _\epsilon [A]$ of the $\theta$-vacuum  proves to be manifestly covariant with respect to "large" $SU(2)$ gauge transformations, as Eq. (\ref {period1}) runs.

Discussing Eq. (\ref {period1}), note that  the raising operator $T_1$ \cite{Cheng}, (\ref {rise}), can be represented explicitly in the shape \cite{Pervush1}
\bea \label{R}
T_1= \exp (\frac {d}{dX[A]})= \exp \left\{\left[\int d^3x B^2 \frac {g^2}{16\pi^2}\right]^{-1} \int d^3x B_i^a \frac {\delta}{\delta A_i^a}\right\}. \eea
This look for $T_1$ follows immediately from the interpretation \cite{Cheng} of the $\theta$-angle as a pseudomomentum operator having the common system of eigenfunctions $\{\Psi _\epsilon [A]\}$ with the momentum operator $\nabla _i^{ab}(A)$.

Then the winding number functional $X[A]$ may be interpreted as the (generalized) coordinate canonically conjugate to the (pseudo)momentum $\theta$ \cite{Pervush1}:
\bea \label{com} 
\left[\frac {d}{dX[A]},X[A]\right]=1. 
\eea 
In particular, at $X[A]=1$, one comes to Eq. (\ref {period1}).

\medskip
The case of zero eigenvalues $\epsilon=0$ of  Hamilton operators $\Hat H$ is, indeed, an especial case in quantum mechanics.

In this case the existence of physical solutions is rather an exception to the rule than a rule in the Euclidian instanton model \cite{Bel} where the raising operator $T_1$ and the Hamilton operator $\hat H$ do not commute indeed \cite{Pervush1,Arsen}: 
$$[\hat H,T_1]\neq 0,  \quad [\hat H,[\hat H,T_1]]\neq 0$$ 
and so on \footnote{ Thus the raising  operator $T_1$ and therefore the winding number functional $X[A]$, due to (\ref{R}), are, indeed, \it cyclic variables \rm \cite{Landau1}.  }.

As it was discussed in Refs. \cite{Pervush1,Arsen,Galperin} (and also  in the later papers \cite{LP2,LP1,Pervush2,fund}), at $\epsilon=0$,  the functional of the plane wave type exists,
\be
\label{plan} \Psi _0[A]= \exp \{i(2\pi k +\theta)X[A]\} \quad (k\in {\bf Z}),
\ee 
satisfying simultaneously the conditions (\ref{ve}), (\ref{period1})  and  the Schr\"odinger equation (\ref {Schrod}) for $\epsilon=0$ and the  ``complex'' momentum\rm:
\be 
\label{im} 2\pi k +\theta =2\pi k\pm i 8\pi^2/ g^2.
\ee 
The expression 
\be 
\label{tm} 
P_{\cal N} \equiv 2\pi k +\theta 
\ee 
 is called 
\it the  momentum 
of the topological motion \rm (e.g. in the terminology of the paper \cite{LP1}) or simply \it the
 topological momentum\rm \footnote{Graphically (on the complex plane), the set of ``complex'' momenta $P_{\cal N}$, (\ref{im}), is a discrete set (more exactly, it is swept by the discrete ``real'' part  $2\pi k$ and continious ``imaginary'' part $i8\pi^2/ g^2$.) }. \par 
Thus the topological momentum 
\be 
\label{tm1} 
P_{\cal N}=\pm i 8\pi^2/ g^2 
\ee 
(at $k=0$)
indeed proves to be purely imaginary. \par 
Additionally, the wave function $\Psi_0[A]$, (\ref{plan}), satisfies the duality equation (\ref {dual1}) (as it was noted in Refs. \cite{Pervush1,Arsen}). 

In the operator shape, the duality equation (\ref {dual1}) has the look \cite{Pervush1}
\be \label{dualop}
E~\Psi_0[A]=\pm B~\Psi_0[A]. \ee  
Moreover, the wave function $\Psi_0[A]$ is specified badly at the minus sign before $P_{\cal N}$ in  (\ref{plan}). Just this implies disappearing the Hamilton operator $\hat H$, (\ref{Schrodinger}) (i.e. $\hat H=\epsilon=0$), due to the asymptotic  conditions (\ref{tens.a}). 

\medskip The said implies that it is impossible to give the correct probability description of the instanton $\theta$-vacuum since the Hilbert space of (topologically degenerated) instanton states becomes non-separable in this case. 

From the quantum-mechanical point of view, the existence of purely imaginary values (\ref{tm1}) of the topological momentum $P_{\cal N}$ in the Euclidian instanton model \cite{Bel} means the possibility of semi-classical paths between "in" and "out" vacua (\ref {cl.vac}) implicating these purely imaginary values of the topological momentum $P_{\cal N}$.

Herewith, as it was discussed in Refs. \cite{LP2,LP1,Pervush2}, appropriate semi-classical transition amplitudes involve the quantum analogue
of an instanton: 
\be \label{wave} \exp (iW [A_{\rm instanton}]=\Psi _0[A=L_{\rm out}]\times \Psi^* _0[A=L_{\rm in}]= \exp(-\frac {8\pi ^2}{g^2}[n_{\rm out}-n_{\rm in}]).
\ee 
It is easy to understand that these "motions" of instantons between "in" and "out" vacua implicating purely imaginary values (\ref{tm1}) of the topological momentum $P_{\cal N}$ cannot be referred to the class of physical motions (and rather to the class of ``ghost'' modes), and below we shall consider an important argument confirming this fact. 




\medskip
On the other hand, the conditions (\ref{ve}), (\ref{period1})  and  the Schr\"odinger equation (\ref {Schrod}) for $\epsilon=0$ can be satisfied also at real values of the $\theta$-angle, as it was discussed in  Ref. \cite{Galperin}.

Really, at acting by the operator ${\hat H}(i\delta/\delta A, A)$ on the wave function  $\Psi_0[A]$, the factor $2\pi k+\theta $ is not important at achieving the result (\ref {Schrod}).

Thus real values of the $\theta$-angle as well as  imaginary ones prove to be relevant for describing the $\theta$-vacuum in the spirit \cite{Callan1}.
In particular, following \cite{Galperin}, one can consider the  real topological momentum
\be \label{tm1real}
P_{\cal N}^{\rm R}= \pm 8\pi^2/ g^2 
\ee
at $k=0$
by analogy with  (\ref{tm1}). 

The said allows to introduce the complex $\theta$-angle:
\be \label{ctheta}
\theta = \theta_1+ i\theta_2, \ee
which imaginary part $\theta_2$ just corresponds to the purely imaginary topological momenta $P_{\cal N}$ \cite{LP1,Pervush1, Pervush2,Arsen}, (\ref {tm1}), just at $k=0$.
Its real part $\theta_1$ may be assumed to vary in the interval $[-\pi,\pi]$ \cite{Pervush1}.

Thus the complex $\theta$-angle (\ref{ctheta}) may be considered as a "compromise variant" united real $\theta$-angles $\theta_1$ \cite{Cheng, Galperin} and purely imaginary $\theta_2$  ones \cite{Pervush1}, given via (\ref{im}), in the Euclidian instanton YM model \cite{Bel}.

\medskip 
Nevertheless, imaginary values $\theta_2$ of the $\theta$-angle and appropriate imaginary values (\ref{tm1}) (at $k=0$) of the topological momentum $P_{\cal N}$, referring to the unphysical spectrum, present always in the Euclidian instanton model \cite {Bel}.

In the paper \cite{Arsen}  this statement was referred to as the so-called \it no-go theorem\rm. 

The results got can be also treated  \cite{Pervush1} as the presence of unphysical solutions to the  Schr\"odinger equation (\ref {Schrod}) at the application of ordinary quantization methods to a topologically nontrivial theory. \par 

\medskip At studying  (\ref{tm1real}) it becomes obvious that such real topological momentum $P_{\cal N}^{\rm R}$ vanishes (i.e. the instanton YM configuration stops) in the limit $g\to\infty$ for the YM coupling constant $g$. It is just the infrared QCD confinement limit as it is understood in modern physic.  In the terminology \cite{Pervush1} this case is referred to as {\it infrared catastrophe}. 

Note that purely imaginary (and thus space-like and unphysical) values (\ref{im}) have no relation to the infrared catastrophe. 

\medskip
There may be shown, repeating the arguments \cite{Cheng}, that    tunnelling transition amplitudes of the (\ref {am}) type may be rewritten  in terms of the 
$\theta$-vacuum \cite{Pervush1, Arsen,Galperin,Callan1}  as \cite{Bel,Cheng} 
\be
\label{ta} 
< \theta' \vert e^ {-iHt}\vert \theta >_J= \delta (\theta'-\theta)I_J(\theta) \ee 
with an unknown function $I_J(\theta)$. \par 
Because of (\ref{rise}), (\ref{decomposition}), 
we can write down: 
\bea 
\label{summa} 
< \theta' \vert e^ {-iHt}\vert \theta >_J &= &
\sum \sb {m,n} e^ {im\theta'}e^ {-in\theta} <m\vert e^ {-iHt}\vert n>_J\nonumber \\ && =\sum \sb {m,n} e^ {-i(n-m)\theta}e^ {im(\theta'-\theta)}\times\nonumber \\ && \times\int [dA]\sb {n-m} \exp \{-i\int ({\cal L}+JA)d^4x\}, 
\eea 
with $J$ being the appropriate non-Abelian current.

One can cast (\ref{summa})  in the standard form by  substituting $n-m \to \nu$;
then upon doing the $m$-summation one gets 
\bea \label{IJ} 
I_J(\theta)&= &\sum \sb {\nu} e^ {-i\nu\theta}\int [dA]\sb {\nu} \exp \left\{-i\int  ({\cal L}+JA)d^4x\right \} \nonumber \\ && =\sum \sb {\nu}\int [dA]\sb {\nu} \exp \left\{-i\int ({\cal L}_{\rm eff}+JA)d^4x\right\},  \eea 
 where, utilising the expression (\ref{Ch-S})  for the Chern-Simon functional, we introduce the 
\it effective YM Lagrangian\rm: 
\be \label{eL}
{\cal L}_{\rm eff} = {\cal L}+ \frac{g^2\theta}{16\pi^2} ~{\rm tr}~ ( F_{\mu \nu}^a \tilde F^{\mu \nu}). 
\ee 
The well-known $\theta$-term appears in the latter expression. 

It determines the contribution
of the natural $SU(2)$ topology 
\be \label{sph} \pi_3 (SU(2))=\pi_3 S^3 = \bf Z \ee
and instanton $\theta$-vacuum to the Euclidian YM Lagrangian formalism. 

It may be  treated as an effective YM selfinteraction with the coupling constant $g^2\theta$. 

The principal shortcoming of Eq.
(\ref{eL}) is its actual dependence on the topological momentum $P_{\cal N}$: (\ref{tm}), (\ref {tm1}).  This  violates the P and CP symmetries \rm in the instanton YM theory \cite{Bel}, the so-called instanton CP-problem arises. \par
The manifest CP-covariance of the effective Lagrangian (\ref{eL}) is accompanied by the bad ultraviolet behaviour of the  Euclidian instanton theory  \cite{Bel} (such behaviour always testifies about a "bad physics").

Moreover (and this fact was pointed out, for instance, in Ref. \cite{Galperin}), the effective instanton Lagrangian (\ref{eL}) is not a Lorentz invariant since the $\theta$-term entering this Lagrangian contains, indeed, the $BE$ product (satisfying the duality condition (\ref {dualop})) that is manifestly Lorentz covariant. 

This implies automatically the Poincare covariance of the Lagrangian (\ref{eL}) due to the natural embedding of the (general) Lorentz group in the Poincare one.

Herewith the Lorentz covariance \cite{Galperin} of the effective instanton Lagrangian (\ref{eL}) supplements its CP-covariance \cite{Cheng}.

Simultaneously, the $\theta$-term in (\ref{eL})  is manifestly gauge invariant as that can be represented in the shape \cite{Al.S.}
$$  \sim \theta \int d^4x <F_{\mu \nu}, F_{\alpha\beta}> \epsilon^{\mu \nu 
\alpha\beta }.$$
Meanwhile, the any product $<F_{\mu \nu}, F_{\alpha\beta}> $ is manifestly gauge invariant \cite{Cheng}.

\medskip
The definite analogy between the $\theta$-term in the effective instanton Lagrangian (\ref{eL}),  involving  imaginary values (\ref{tm1}) of the topological momentum $P_{\cal N}$ (that corresponds to the imaginary part $\theta_2 $ of the $\theta$-angle), and the action functional (\ref{rot}), describing correctly collective solid rotations inside the Minkowskian BPS monopole vacuum suffered the Dirac fundamental quantization \cite{Dir} can be observed.

Really, at identifying the ends of discussed semi-classical paths \footnote{As it was discussed in Ref. \cite{Postn4}, such identifying may be performed by going over to the compact Euclidian space ${\bf R}^4\bigcup \{\infty\}$ from the $E_4$ one.

In  this case the asymptotic (\ref{tens.a}), (\ref{EB}) \cite{Cheng,Lenz} for the YM tension tensor $F_{\mu\nu}^{a}$ is mathematically equivalent to continuing this tensor with its zero value in the point $\infty$. 

On the other hand, ${\bf R}^4\bigcup \{\infty\}\simeq S^4$ (via the stereographical projection has been performed \cite{Postn4}).}, one gets closed (three-dimensional) trajectories of infinite "radiuses".

In  this case semi-classical paths  between "in" and "out" instanton vacua (\ref{cl.vac}) come to circular motions in $S^4$ some of which proceed with imaginary (topological) momenta (\ref{tm1}). 

Just these ("real" and "imaginary") closed trajectories are described correctly by the $\theta$-term in the effective instanton Lagrangian (\ref{eL}).

On the other hand, as it follows from (\ref{eL}) \cite{Cheng}, at $\epsilon=0$ (that corresponds to instanton field configurations with ${\bf E}={\bf B}=0$ at the spatial infinity \cite{Cheng,Lenz}), only the $\theta$-term is actual in (\ref{eL}).

On the face of it, the $\theta$-term in  (\ref{eL}) would give additional nonzero contributions in the energy integral of the instanton model \cite{Bel} (in comparison with $\epsilon=0$ coming indeed from ${\cal L}$).

But two essential objections may be brought at once against the latter assertion. 

Firstly, some of abovementioned circular three-dimensional motions (that are, geometrically, the spheres $S^3$ of infinite diameters) are accompanied by {\it negative} kinetic energies \cite{Galperin} $\propto P_{\cal N}^2$ (with $ P_{\cal N}$ now being purely imaginary).

Secondly, the circular three-dimensional motions with {\it positive} kinetic energies (corresponding to real values of $P_{\cal N}$ and the real part $\theta_1 $ of the $\theta$-angle) are, indeed, fictive since the $\theta$-term in the instanton YM effective Lagrangian (\ref{eL}) \cite{Cheng} does not alter  the YM  equations of motions \footnote{ These equations of motions follow immediately from the action functional $S_{\rm Eucl}$ \cite{Cheng}, (\ref {aYM}), of the instanton YM model \cite{Bel}. }
$$   D_\mu F^{\mu\nu}=0     $$
(this fact was pointed out, for example, in Ref. \cite{Al.S.}).

\medskip
The important conclusion may be drawn from our discussion.

It is out of the question the {\it physical} Jksephson effect \cite{Pervush3} in the Euclidian instanton YM model \cite{Bel} (unlhke that taking place in the M)nkowskhan Higgs modal with vacuum BPS monopolas quantized by Dirac \cite{Dir},  il QED$W{(1+1)}$  \cite{Ilieva1} or in the liqui` heliue II at best theory Lcite{Halatnikov}).

In three mentioned models the Josephson effect \cite{Pervush3} taking place is just a physical effect. 

For instance, in the Minkowskian Higgs model with vacuum BPS monopoles quantized by Dirac, the appropriate  action functional (\ref{rot}) \cite{LP2,LP1, Pervush2,David2} and the purely real energy-momentum spectrum $P_N$, (\ref{pin}), reading immediately from the action functional (\ref{rot}), determine the physical nature of the Josephson effect taking place. 

In this case \cite{Pervush3}, the Hamilton operator
\be\label{Hamil}
\hat H_1= \frac {P_N^2}{2I}
\ee 
corresponds to the free rotator action functional (\ref{rot}), describing collective solid rotations inside the Minkowskian Higgs BPS monopole vacuum suffered the Dirac fundamental quantization \cite{Dir}.

Then, instead of the Schr${\rm \ddot o}$dinger equation (\ref {Schrodinger}) \cite{Pervush1} inherent in the Euclidian instanton YM model \cite{Bel},  the Schr${\rm \ddot o}$dinger equation 
\be \label{srph}
\hat H_1\Psi(N)=\epsilon_\theta \Psi(N),
\ee
with $\epsilon_\theta $ being the kinetic (rotary) energy, don't vanishing until $\theta \neq 0$, corresponding  to the free rotator action functional (\ref{rot}) represented in the shape \cite{Pervush2, Pervush3,Arsen}
\be
 \label{rotat}
 W_{\rm rot}= \int dt  \frac{ P_N^2 (t)}{2I}
 \ee
takes place in the mentioned Minkowskian Higgs model. 

And moreover, as it is easy to see, herewith the topological momentum $P_N$, (\ref{pin}), and the topological dynamical variable $N(t)$, given via Eq. (\ref {winding num.}), prove to be canonically conjugated values \cite{Pervush3}:
\be
 \label{conjug}
i[P_N, N(t)]=\hbar.
\ee
Further, unlike the Euclidian instanton YM model \cite{Bel}, where the appropriate Hamilton operator $\hat H$, (\ref {Schrod}), does not commute with the raising operator $T_1$ given via (\ref {R}) (this is equivalent to nonzero commutators also between $\hat H$ and the topological momentum $P_{\cal N}$), now the Hamilton operator $\hat H_1$, (\ref {Hamil}), contains explicitly the topological momentum $P_{ N}$, (\ref {pin}). Thus the values $P_{ N}$ and $\hat H_1$ commute:
\be
 \label{comt}
[P_{ N}, \hat H_1]=0.
\ee
In  this case it is necessary to replace the raising operator $T_1$: (\ref {period1}), (\ref {R}), proper to the Euclidian instanton YM model \cite{Bel}, with the one \cite{Arsen}
\be \label{R2}
T_{G1}= \exp (i  P_N)= \exp (\frac{d}{d N(t)})
\ee 
in order  that $ [T_{G1},\hat H_1]=0$.

The commutation relation (\ref{comt}) between $ P_{ N}$ and $\hat H_1$ proper to the Minkowskian Higgs model with vacuum BPS monopoles quantized by Dirac \cite{Dir} may be treated also as the (sufficient and necessary) condition that the energy-momentum spectrum (\ref{pin}) is real in that model (this follows from the well-known fact \cite{Landau3} that Hamilton operators are Hermitian and involve real eigenvalues).

\medskip
The  fact that the Hamilton operator $H_1$, (\ref {Hamil}), is squared by the topological momentum $ P_{ N}$ is also remarkable. 

This indicates the manifest relativistic (Poincare) invariance of this Hamilton operator (unlike the relativistic covariant $\theta$-term in the effective Lagrangian ${\cal L}_{\rm eff}$ \cite{Cheng}, (\ref {eL}), inherent in the Euclidian instanton YM model \cite{Bel}). 

This seems to be the way to avoid the CP-problem in the Minkowskian Higgs model with vacuum BPS monopoles quantized by Dirac \cite{Dir}.

\bigskip
Note also  that introducing the complex $\theta$-angle, (\ref{ctheta}), in the Euclidian instanton YM model \cite{Bel} implicates singularities in semi-classical transitions amplitudes (\ref {summa}) between asymptotical "in" and "out" states. 

Now we shall attempt to demonstrate this fact.
The arguments stated in Ref. \cite{Kurr} will help us to do this. \par 
The account of the $\theta$-term in the effective instanton Lagrangian (\ref{eL}) allows to represent any transitions amplitude (\ref {summa}) between vacuum "in" and "out" states in the shape \cite{Bel, Kurr}\footnote{The result below was got in Ref. \cite{Kurr} for the minimal nontrivial topologies $n=\pm1$ (or $N=\vert n\vert=1$), while instantons with $N\geq 2$ give subdominant contributions in the Euclidian  instanton model \cite{Bel}.\par
Moreover, in Ref. \cite{Kurr}, the $\theta$-term was assumed to be 
$ -i N \theta$. Herewith the presence of the $i$ multipliers does not change the equations of motion in the model \cite{Bel}and maintains the gauge invariance of the $\theta$-term. \par
In this case  $\cos \theta 
$ multipliers appear in transitions amplitudes between vacuum "in" and "out" states, as it follows from  general Eq. (\ref {am}) \cite{Cheng} for Euclidian semi-classical transitions amplitudes.}
\be 
\label{Gdg}
\Gamma \propto \exp(-8\pi^2/g^2) \cos \theta 
\ee
For complex $\theta$-s, as it is well known \cite{Schabad}, 
\be 
\label{cocos} 
\cos\theta = \frac{e^{i\theta }+ e^{-i\theta }} 2
\ee
As it was argued in \cite{Schabad}, this function of $\theta_1 +i\theta_2 $ acquires values can be any amount large if $\theta_1 ={\rm Re}~\theta =\pm \pi$, i.e., indeed, it would be assumed that $\theta_1 \in ]-\pi,\pi[$, instead of $\theta_1 \in [-\pi,\pi]$ as it was done in Ref. \cite{Pervush1}.
\medskip

The pointed singularities of the "complex" $\cos\theta $ \cite{Schabad} influence also the energy contribution $\epsilon\neq 0$ from the $\theta$-term in the effective instanton Lagrangian ${\cal L}_{\rm eff}$  \cite{Cheng} (\ref{eL}).

Indeed, this contribution \cite{Pervush3}
$$\epsilon_\theta \sim V\cos \theta_1; \quad V= 4\pi R^3/3 $$
(with $\theta =\theta_1 +i\theta_2 $ given via (\ref{ctheta})), even without that infinite in the limit $V=R^3\to\infty$, becomes additionally singular at $\theta_1 =\pm \pi$ \cite{Schabad}.

\end{appendix}
 
 \end{document}